\documentclass{jpp}
\usepackage{graphicx}
\usepackage[export]{adjustbox}
\usepackage{comment, color}

\usepackage[utf8]{inputenc}
\usepackage[T1]{fontenc}
\usepackage{amsmath}

\newcommand{\vect}[1]{\mbox{\boldmath $#1$}}

\newcommand{\changed}[1]{{#1}}
\newcommand{\Gsign}{ s_G }
\newcommand{\psisign}{ s_{\psi} }
\newcommand{\boozertor}{\varphi}

\shorttitle{Direct construction of optimized stellarator shapes. II.}
\shortauthor{M. Landreman, W. Sengupta, and G. G. Plunk}

\title{Direct construction of optimized stellarator shapes. II. Numerical quasisymmetric solutions}

\author{Matt Landreman\aff{1}
\corresp{\email{mattland@umd.edu}},
 Wrick Sengupta\aff{2}
 \and Gabriel G Plunk\aff{3}}

\affiliation{\aff{1}Institute for Research in Electronics and Applied Physics, University of Maryland, College Park MD 20742, USA
\aff{2}Courant Institute of Mathematical Sciences, New York University, New York NY 10012, USA
\aff{3}Max Planck Institute for Plasma Physics, Greifswald, Germany}

\begin{document}

\maketitle

\begin{abstract}

Quasisymmetric stellarators are appealing intellectually and as fusion reactor candidates since the guiding center particle trajectories and neoclassical transport are isomorphic to those in a tokamak, 
\changed{implying good confinement.}
Previously, quasisymmetric magnetic fields have been identified by applying black-box optimization algorithms to minimize symmetry-breaking Fourier modes of the field strength $B$. Here instead we directly construct magnetic fields in cylindrical coordinates that are quasisymmetric to leading order in distance from the magnetic axis, without using optimization. The method involves solution of a 1-dimensional nonlinear ordinary differential equation, originally derived by Garren and Boozer [\emph{Phys. Fluids} B {\bf 3}, 2805 (1991)].
We demonstrate the 
usefulness and accuracy
of this optimization-free approach by providing the results of this construction as input to the
codes VMEC and BOOZ\_XFORM, confirming the purity and scaling of the magnetic spectrum. The space of magnetic fields
that are quasisymmetric to this order is parameterized by
the magnetic axis shape
along with three other real numbers,
one of which reflects the on-axis toroidal current density, and another
one of which is zero for stellarator symmetry.
The method here could be used to generate good initial conditions for conventional optimization, and its speed enables exhaustive searches of parameter space.

\end{abstract}

\section{Introduction}

Toroidal magnetic fields can posses a remarkable hidden symmetry, called quasi-symmetry, in which the field strength $B=|\vect{B}|$ is independent of a particular coordinate (``Boozer angle'') even though the magnetic field vector $\vect{B}$ is not \citep{Boozer83, NuhrenbergZille, HelanderReview}. Since the Lagrangian
for guiding-center particle motion in Boozer coordinates varies on magnetic surfaces only through $B$, a symmetry direction in $B$ implies
that guiding-center trajectories behave as if the magnetic field had a true symmetry direction, and the conserved quantity that follows from Noether's theorem implies that particle trajectories are confined. In contrast, magnetic fields without continuous symmetry generally have unconfined guiding-center trajectories. 
\changed{(Quasisymmetry is sufficient but not necessary for guiding center confinement \citep{CaryShasharinaPRL}.) }
Plasmas confined by quasisymmetric 
magnetic fields are also predicted to have temperature screening of impurities and to \changed{allow} larger flows, which may lead to improved stability. For these reasons, quasisymmetric magnetic fields are  interesting both for fusion energy and on basic physics grounds.

A number of quasisymmetric magnetic configurations have been identified to date
 \citep{NuhrenbergZille, NuhrenbergQA, HSX, Garabedian, NCSX, KuBoozerQHS, ESTELL,PlunkHelander,Henneberg,ROSE}. In all of these cases except the work of \cite{PlunkHelander}, the quasisymmetric configurations have been found using optimization, by minimizing the amplitudes of symmetry-breaking Fourier modes of $B$. The optimization algorithms used have been ``off the shelf'' algorithms that can be applied to minimizing any function and do not exploit  information about the underlying physical system. While this approach has proven successful, it does have a number of shortcomings. Little insight is provided as to the form and dimensionality of the landscape of solutions. As the results of the optimization depend on the initial guess and on manually chosen weight parameters, there is no guarantee that all interesting solutions have been found. Optimization is also computationally demanding, requiring many 3D equilibrium calculations.

A complementary approach to finding quasisymmetric geometries, developed by \cite{GB1}, is to directly construct the geometry from the relevant equations, with no need then for optimization. Expanding in small distance $r$ from the magnetic axis (that is, large aspect ratio), Garren \& Boozer derived equations for quasisymmetry to first and second order in $r$. Their work is perhaps best known for the result that the number of equations exceeds the number of unknowns at third order, 
\changed{so quasisymmetry may be achieved on one surface but not throughout a volume. However} the useful constructive procedure at lower order has not been fully exploited as a tool to generate quasisymmetric shapes, which can be useful as initial conditions for conventional optimization, and to understand the landscape of quasisymmetric shapes. The goal of this paper is to reinvigorate this development. 

In an accompanying Paper I \citep{PaperI}, we derived two ways to generate a shape in standard cylindrical coordinates with prescribed $B$ using the Garren-Boozer framework, summarized in sections \ref{sec:conversion}  and \ref{sec:conversion3} below. In the present paper, we develop the optimization-free approach to constructing quasisymmetric geometries in several ways. In section 3, we present a new spectrally accurate algorithm for solving the equation for quasisymmetry to first order in $r$.  Using several methods for converting the results to standard cylindrical coordinates, explained in section 4, we present in section 5 examples of quasi-axisymmetric and quasi-helically symmetric equilibria obtained without optimization. 
\changed{(Quasi-poloidally symmetric configurations cannot be generated using this approach since this symmetry is impossible near the axis.)
For each of these configurations,}
we use the codes VMEC \citep{VMEC1983, VMEC1986} and BOOZ\_XFORM \citep{Sanchez} to compute the spectrum of $B$, confirming that the symmetry-breaking harmonics are small and that they scale as expected. One family of equilibria we consider (section 5.3) possesses quasisymmetry but not stellarator symmetry, which may be desirable for obtaining significant intrinsic rotation. We discuss and conclude in section 6. A proof that a unique solution to the first-order quasisymmetry equation exists despite the nonlinearity of the problem is given in the appendix.

The approach here allows quasisymmetric flux surface shapes
to be computed in $<1$ millisecond on a laptop. This timescale is at least 4 orders of magnitude faster than 
a typical equilibrium calculation with VMEC, much less an optimization in which VMEC is iterated to find quasisymmetric equilibria. 
Our approach can therefore be used for extensive searches of parameter space, potentially enabling an identification of all possible quasisymmetric plasma shapes,
at least in the vicinity of the magnetic axis.
Also, this ``direct construction'' approach  makes clear how many degrees of freedom are
available in the space of quasisymmetric magnetic fields, giving insight into the landscape of solutions.

 
\section{System of equations}
\label{sec:equations}

Here we summarize the  equations relevant to first-order quasisymmetry
derived in \cite{GB1}.
The position vector can be written in flux coordinates as
\begin{equation}
\vect{r}(r,\theta,\boozertor)=\vect{r}_0(\boozertor) + r X_1(\theta,\boozertor) \vect{n}(\boozertor) + r Y_1(\theta,\boozertor) \vect{b}(\boozertor) + O(r^2),
\label{eq:GBr}
\end{equation}
where $\vect{r}_0$ is the position of the magnetic axis,
$(\theta,\boozertor)$ are the poloidal and toroidal Boozer angles, $r=\sqrt{2|\psi|/B_0}$ is an effective minor radius that labels flux surfaces,  $2\pi\psi$ is the toroidal flux, and $B_0$ is the magnetic field strength along the axis, which must be a constant due to quasisymmetry. The unit normal vector $\vect{n}$ and unit binormal $\vect{b}$ are defined in terms of the magnetic axis shape by the Frenet-Serret relations
\begin{align}
d\vect{t}/d\ell &= \kappa \vect{n}, \\
d\vect{n}/d\ell &= -\kappa \vect{t} + \tau \vect{b}, \nonumber \\
d\vect{b}/d\ell &= -\tau \vect{n}, \nonumber
\end{align}
where $\vect{t}(\boozertor) = d\vect{r}_0/d\ell$ is the unit tangent vector, $\vect{t}\cdot\vect{n}\times\vect{b}=1$, $\ell$ denotes arclength, $\kappa(\boozertor)$ is the 
curvature, and $\tau(\boozertor)$ is the torsion. (Garren and Boozer use the opposite sign convention for torsion.)
The $\vect{t}$ component in (\ref{eq:GBr}) at $O(r)$ can be shown to vanish.
To first order in the distance from the magnetic axis, the flux surface shape is described by
\begin{align}
X_1 = X_{1s}(\boozertor)\sin\theta + X_{1c}(\boozertor)\cos\theta,
\hspace{0.5in}
Y_1 = Y_{1s}(\boozertor)\sin\theta + Y_{1c}(\boozertor)\cos\theta,
\end{align}
and the magnetic field strength satisfies
\begin{equation}
B(r,\theta,\boozertor)=B_0 + r \left[ B_{1s}(\boozertor)\sin\theta + B_{1c}(\boozertor)\cos\theta\right] + O(r^2),
\label{eq:Bexpansion}
\end{equation}
where
\begin{align}
X_{1s}(\boozertor) = B_{1s}(\boozertor) / [B_0 \kappa(\boozertor)],
\hspace{0.5in}
X_{1c}(\boozertor) = B_{1c}(\boozertor) / [B_0 \kappa(\boozertor)].
\end{align}
In the case of quasisymmetry, we can choose the origin of the $\theta$ coordinate so $B=B_0 + r \bar\eta B_0\cos(\theta-N\boozertor)+O(r^2)$ for some constant $\bar\eta$ and fixed integer $N$, with quasi-axisymmetry defined by $N=0$ and quasi-helical symmetry defined by $N \ne 0$.
Then $B_{1s}= \bar\eta B_0 \sin(N\boozertor)$ and $B_{1c}=\bar\eta B_0 \cos(N\boozertor)$.
Furthermore,
\begin{align}
\label{eq:W_to_Y}
Y_{1s} = (\Gsign \psisign \kappa /\bar\eta)\left[\sigma \sin(N\boozertor) + \cos(N\boozertor)\right], \hspace{0.3in}
Y_{1c} = (\Gsign \psisign \kappa /\bar\eta)\left[\sigma \cos(N\boozertor) - \sin(N\boozertor)\right],
\end{align}
where $\Gsign=\pm 1$ is positive (negative) if $\vect{B}$ points towards increasing (decreasing) $\boozertor$, $\psisign = \mathrm{sign}(\psi)=\pm 1$, and the periodic function $\sigma(\boozertor)$ satisfies
the Riccati-type equation
\begin{align}
\frac{d\sigma}{d\boozertor}
+(\iota-N) \left[ 
\frac{\bar\eta^4}{\kappa^4}
+1+\sigma^2\right]
- \frac{2G_0\bar\eta^2}{B_0 \kappa^2} \left[ \frac{I_2}
{B_0} -\psisign\tau\right] 
=0.
\label{eq:QA}
\end{align}
Here, $\iota$ is the rotational transform on axis; 
$G_0$ is the on-axis value of $G(r)$, the poloidal
current outside the flux surface times $\mu_0/(2\pi)$;
and $I_2$ is the leading coefficient in $I(r) = r^2 I_2 + O(r^4)$, the toroidal
current inside the flux surface times $\mu_0/(2\pi)$. The functions $G(r)$ and $I(r)$ here are those appearing in the Boozer coordinate representation
\begin{align}
\vect{B} = \beta(r,\theta,\boozertor)\nabla r
+ I(r) \nabla\theta + G(r) \nabla\boozertor.
\end{align}
Using Eq (2.18) in Paper I,
eq (\ref{eq:QA}) can be written in terms of the standard toroidal angle $\phi$ (the azimuthal angle in cylindrical coordinates $(R,\phi,z)$)
as
\begin{equation}
\frac{d\sigma}{d\boozertor}
 = \frac{|G_0|}{\ell' B_0} \frac{d\sigma}{d\phi},
 \label{eq:zeta_to_phi}
\end{equation}
where
\begin{equation}
\ell' = d\ell/d\phi = \sqrt{R_0^2 + (R'_0)^2 + (z'_0)^2},
\end{equation}
the magnetic axis has cylindrical coordinates $R_0(\phi)$ and $z_0(\phi)$, and primes denote $d/d\phi$.
Also, $G_0$ can be related to the magnetic axis shape by $
G_0=\Gsign B_0 L/(2\pi)$
where $L=\int_0^{2\pi}d\phi\; \ell'$ is the length of the axis.
The above equations all apply even if the plasma pressure is nonzero, although the pressure turns out not to appear in these expressions to this order.

As discussed in section 5.2 of Paper I, $N$ can be determined directly from the axis shape. The integer $N$ is
the number of times the normal vector $\vect{n}$ rotates poloidally around the magnetic axis as the axis is traversed once toroidally.
We can determine $N$ this way because the vector at each $\boozertor$ pointing from the magnetic axis to the maximum-$B$ contour on a flux surface $r$ is $\vect{n} r |\bar{\eta}| / \kappa + \vect{b} r Y_1$, which has a positive projection onto $\vect{n}$ at every $\boozertor$. Hence these two vectors are always within 90 degrees of each other, and so the $B$ contours loop around the magnetic axis the same number of times $\vect{n}$ does so.

Stellarators, whether quasisymmetric or not, typically are designed to possess stellarator \changed{symmetry,
since this symmetry reduces the dimensionality of the parameter space for optimization, reduces the computational cost of many calculations, and typically reduces the number of unique coil shapes. Stellarator symmetry} corresponds to $R(-\theta,\phi) = R(\theta,\phi)$,
$z(-\theta,-\phi)=-z(\theta,\phi)$, and $B(-\theta,-\phi)=B(\theta,\phi)$. For a magnetic field described by  (\ref{eq:GBr})-(\ref{eq:QA}) to be stellarator-symmetric, the magnetic axis shape must be stellarator-symmetric, and $\sigma(\boozertor)$ should be odd. 

As proved in the appendix, even though (\ref{eq:QA}) is nonlinear in $\sigma$,
this equation can be posed in such a way that there is guaranteed to be
precisely one solution. Specifically, given well-behaved $\kappa(\boozertor)$, $\tau(\boozertor)$, $I_2/B_0$, $G_0/B_0$,  
$\bar\eta$,
and an initial condition $\sigma(0)$, 
there is precisely one solution pair $\{\iota, \, \sigma\}$ such that $\sigma(\boozertor)$ is periodic. As a result, for any magnetic axis shape with nonvanishing curvature,
there are an infinite number of magnetic fields in the vicinity of that axis which are consistent with quasisymmetry to first order in $r$. The possible magnetic fields are parameterized by three numbers: $I_2/B_0$, $\bar\eta$, and $\sigma(0)$.
If the current density vanishes on axis, which is common in stellarators even at finite plasma pressure since the bootstrap current density vanishes on axis, $I_2 = 0$. Furthermore, for stellarator-symmetric fields, $\sigma(0)=0$. Therefore, in practice usually only one of the three scalar input parameters is free. While every magnetic axis shape with nonvanishing curvature admits an infinite number of quasisymmetric fields, for many axis shapes \changed{it is found numerically that} the elongation of the surrounding flux surfaces reaches enormous values (tens, hundreds, or thousands), making the shape uninteresting.


\section{Numerical method}
\label{sec:numerical}

For a practical solution of (\ref{eq:QA}) we consider the inputs to be
 $\bar\eta$, $I_2/B_0$,
 $\sigma(0)$, and the shape of the magnetic axis $\{R(\phi),\,z(\phi)\}$. The outputs are $\sigma(\phi)$ and $\iota$.
Given the inputs, we solve (\ref{eq:QA}) with (\ref{eq:zeta_to_phi}) for $\sigma(\phi)$ using Newton iteration with a pseudo-spectral collocation discretization.
A uniform grid of \changed{$N_\phi$} points,
$\phi_j = (j-1) 2\pi/(N_\phi n_{fp})$ where $j = 1\ldots N_\phi$,
is defined on $[0,\, 2\pi/n_{fp})$, where $n_{fp}$ is the number
of identical field periods. The vector of unknowns
is taken to be $[\iota, \, \sigma_2, \ldots, \sigma_{N_\phi}]^T$
where $\sigma_j=\sigma(\phi_j)$, so there are $N_\phi$ unknowns.
There is no need to include $\sigma(\phi_1)=\sigma(0)$ as an unknown since it is a prescribed input.
A system of $N_\phi$ equations is obtained by imposing (\ref{eq:QA}) at all $\phi_j$. The $d\sigma/d\phi$ 
derivative is discretized using the Fourier pseudo-spectral differentiation matrix \citep{DMSuite}.
Newton iteration proceeds by solving linear systems involving the Jacobian matrix
$[\partial \vect{R}/\partial\iota, \, \partial \vect{R}/\partial \sigma_2,\, \ldots,\, \partial \vect{R}/\partial \sigma_{N_\phi}]$,
where $\vect{R}$ is the residual vector.
It is straightforward to analytically evaluate the derivatives in the Jacobian in terms of the differentiation matrix.
For the examples shown below, the residual $L^2$ norm is reduced by $15$ orders of magnitude in $\le 5$ Newton iterations.
The numerical solution is extremely robust: in parameter
scans to date we have not observed any examples in which the
Newton iteration fails to converge.

\begin{figure}
  \centering
  \includegraphics[width=3in]{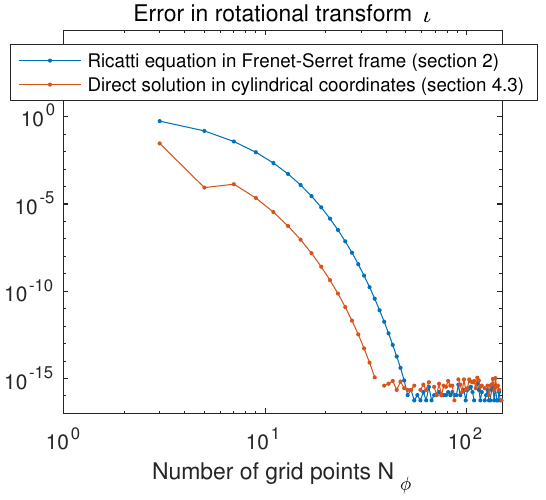}
  \caption{The algorithm of section \ref{sec:numerical} allows
  the equations of section \ref{sec:equations} or section \ref{sec:conversion3} to be solved to machine precision with a modest number of grid points $N_\phi$. The equations of these two sections yield results that are identical to machine precision (for sufficient $N_\phi$) since the equations are equivalent, as proved in Paper I.}
\label{fig:convergence}
\end{figure}

Figure \ref{fig:convergence} demonstrates the convergence
of the rotational transform computed by this method
as the number of grid points increases, for the example of section \ref{sec:QA}. As expected, the convergence is spectral, and $\iota$ can be computed to 15 digits of precision with $N_\phi\sim 50$. In the figure, the `true' value of $\iota$ is taken to be the result for $N_\phi=149$.

\section{Conversion to cylindrical coordinates}

To take advantage of stellarator physics codes that accept VMEC equilibrium files,
such as the STELLOPT optimization suite,
we wish to transform the solutions from the Frenet-Serret frame to the VMEC input representation. In this representation, the plasma boundary is expressed as
a Fourier expansion of the cylindrical coordinates $R(\theta,\phi)$ and $z(\theta,\phi)$,
where $\theta$ can be any poloidal angle. 
Here, we will continue to let $\theta$ be the poloidal Boozer angle.
We can compute $R(\theta,\phi)$ and $z(\theta,\phi)$ from the asymptotic large-aspect-ratio solution in several ways,
described in the following subsections.
The first three approaches have the common feature that an expansion in $r$ for the surface shape is evaluated at a finite value of $r$.

\subsection{First-order method}
\label{sec:conversion}

In one approach, the solution in the Frenet-Serret frame is  transformed to cylindrical coordinates using the method detailed in Paper I,
summarized here.
The position vector is expressed as
\begin{align}
\vect{r} = \hat{\vect{r}}_0(\phi) + r \left[R_1(\theta,\phi) \vect{e}_R(\phi) + z_1(\theta,\phi) \vect{e}_z\right] + O(r^2),
\end{align}
where $\hat{\vect{r}}_0(\phi)$ = $R_0(\phi)\vect{e}_R(\phi) + z_0(\phi) \vect{e}_z$,
and equated to (\ref{eq:GBr}). The $\vect{n}$ and $\vect{b}$ components of the result
give (to leading order in $r$)
\begin{equation}
\begin{pmatrix} X_1 \\ Y_1 \end{pmatrix}
= 
\begin{pmatrix}
n_R & n_z \\
b_R & b_z
\end{pmatrix}
\begin{pmatrix} R_1 \\ z_1 \end{pmatrix},
\label{eq:transformation1}
\end{equation}
where $n_R = \vect{n}\cdot\vect{e}_R$, $b_z = \vect{b}\cdot\vect{e}_z$, etc, 
and this matrix equation can be inverted to give
\begin{equation}
\begin{pmatrix} R_1 \\ z_1 \end{pmatrix}
= 
\frac{\ell'}{R_0}\begin{pmatrix}
-b_z & n_z \\
b_R & -n_R
\end{pmatrix}
\begin{pmatrix} X_1 \\ Y_1 \end{pmatrix}.
\label{eq:transformation}
\end{equation}
Since $X_1$ and $Y_1$ each have $\sin\theta$ and $\cos\theta$ components,
the same is true of $R_1$ and $z_1$, so the flux surfaces are ellipses
in the $R$-$z$ plane. Eq (\ref{eq:transformation}) can be applied at each $\phi$ to
both the $\sin\theta$ and $\cos\theta$ components. Then given
any choice for $r$, a finite-aspect-ratio magnetic surface can be formed
in cylindrical coordinates from $R(\theta,\phi) = R_0(\phi) + r R_1(\theta,\phi)$
and $z(\theta,\phi) = z_0(\phi) + r z_1(\theta,\phi)$.
Note that the transformation (\ref{eq:transformation}) represents only the leading order behavior in an expansion in $r$, so for finite $r$ the flux surface geometry will depart somewhat from (\ref{eq:GBr}), meaning cross-sections of the boundary surface normal to the magnetic
axis will no longer be perfect ellipses.

\subsection{Alternative method}
\label{sec:conversion2}

The method of the previous subsection results in a flux surface shape
that is consistent with (\ref{eq:GBr}) to $O(r)$. Alternatively, 
one can compute the surface defined by the terms through $O(r)$ in (\ref{eq:GBr}) as follows. First a positive value of $r$ is chosen. Given uniform grids in $\theta$ and $\phi$, a tensor product grid is formed. For each point $(\theta,\phi)$ in this tensor product grid, a 1D root finding problem is solved to find $\boozertor$ such that the position vector (\ref{eq:GBr}) has toroidal angle $\phi$. In this way, $R$ and $z$ are obtained on the $(\theta,\phi)$ grid, and so they can be Fourier transformed to provide input to VMEC. To $O(r)$ the resulting surface is identical to the surface constructed in the previous subsection, but $O(r^2)$ differences are present.
This second method ensures cross-sections of the flux surfaces perpendicular to the magnetic axis are elliptical, while cross-sections in the $R$-$z$ plane will generally not be elliptical.

For the quasi-axisymmetric examples below, we find the two methods for
converting to cylindrical coordinates yield nearly indistinguishable results, and so there is no need for the extra complexity of the second method.
However, for the quasi-helically symmetric example below, we find the second
method yields smaller symmetry-breaking harmonics by a factor $\sim 2$, and so we will use it in section \ref{sec:QH}. 


\subsection{Direct solution in cylindrical coordinates}
\label{sec:conversion3}

There is also a third approach to computing $R(\theta,\phi)$ and $z(\theta,\phi)$ for first-order quasisymmetric magnetic surface shapes:
directly solving the first-order quasisymmetry equations
in cylindrical coordinates rather than in the Frenet-Serret frame.
The first-order quasisymmetry equations in cylindrical coordinates were
derived in Paper I and are
\begin{align} 
\label{eq:cylindrical1}
R_{1s}  z_{1c} - R_{1c} z_{1s}-\Gsign  \ell' / R_0 &= 0, \\
K_R R_{1c} + K_z z_{1c} - B_{1c} / B_0 &= 0, \\
K_R R_{1s} + K_z z_{1s} - B_{1s} / B_0 &= 0, \\
\iota  V - T &= 0,
\label{eq:cylindrical4}
\end{align}
where
\begin{align}
K_R = \kappa \vect{n}\cdot\vect{e}_R 
\hspace{0.5in}
K_R = \kappa \vect{n}\cdot\vect{e}_z 
\end{align}
\begin{align}
\label{eq:numerator}
T = 
\frac{|G_0|}{ (\ell')^3 B_0}
 &\left[
R_0^2 \left( R_{1c} R'_{1s}-R_{1s}R'_{1c}+z_{1c}z'_{1s}-z_{1s}z'_{1c}\right) \right. \\
&+\left(R_{1c}z_{1s}-R_{1s}z_{1c}\right) 
\left(R'_0 z''_0 
+2R_0 z'_0
-z'_0 R''_0 \right)\nonumber\\
&+\left(z_{1c}z'_{1s}-z_{1s}z'_{1c}\right)\left(R'_0\right)^2
+\left(R_{1c}R'_{1s}-R_{1s}R'_{1c}\right)\left( z'_0 \right)^2  \nonumber \\
&+\left.\left(R_{1s}z'_{1c}-z_{1c}R'_{1s}+z_{1s}R'_{1c}-R_{1c}z'_{1s}\right) R'_0 z'_0\right]
+\frac{2 G_0 I_2}{B_0^2}  \nonumber
\end{align}
and
\begin{align}
\label{eq:denominator}
V= &
\frac{1}{(\ell')^2}
\left[
R_0^2 \left( R_{1c}^2 + R_{1s}^2 + z_{1c}^2 + z_{1s}^2 \right)
+\left(R'_0\right)^2 \left(z_{1c}^2+z_{1s}^2\right) \right.
\\
&\left.  \hspace{0.5in} -2R'_0z'_0\left(R_{1c}z_{1c}+R_{1s}z_{1s}\right)
+\left(z'_0\right)^2\left(R_{1c}^2+R_{1s}^2\right)\right], \nonumber 
\end{align}
and primes denote $d/d\phi$. As proved in Paper I,
these equations are exactly equivalent
to (\ref{eq:QA}) under the transformation (\ref{eq:transformation1})-(\ref{eq:transformation}).

The system (\ref{eq:cylindrical1})-(\ref{eq:cylindrical4})
can be solved with Newton's method using a procedure similar to the one of section \ref{sec:numerical}. The vector of unknowns consists of $R_{1c}$, $R_{1s}$, $z_{1c}$, and $z_{1s}$, each evaluated at $\phi_j$, along with $\iota$. The same number of equations are obtained by imposing (\ref{eq:cylindrical1})-(\ref{eq:cylindrical4}) at each of the $\phi_j$, along with one additional equation corresponding to the initial condition for $\sigma$.
We verified that this direct solution in cylindrical coordinates
(\ref{eq:cylindrical1})-(\ref{eq:cylindrical4}) indeed yields indentical
results to the method of section \ref{sec:conversion}, within discretization error that can be made as small as machine precision, as shown in figure \ref{fig:convergence}.


\subsection{Outward extrapolation using specific coils}
\label{sec:conversion4}

A fourth method for generating
finite-size quasisymmetric plasma shapes from the
high-aspect-ratio theory, which we now describe,
can potentially generate shapes with relatively low aspect ratio that are realizable with reasonable coils,
at least for the limited case of vacuum fields.
In this method, first one of the methods of sections \ref{sec:conversion}-\ref{sec:conversion3} is used to generate a flux surface shape at a high aspect ratio.
Next, a coil design code such as REGCOIL \citep{regcoil} or FOCUS \citep{FOCUS} is used to find coil
shapes that produce this high-aspect-ratio surface,
by minimizing the (squared) magnetic field normal to the surface.
Typically, good flux surfaces will be produced by these coils well outside of the original target surface, and field line tracing can be used to identify a large region filled with good surfaces. If desired, VMEC can be run in free-boundary mode to obtain a representation of the field in this larger region.
This fourth approach results in boundary surface shapes that are not strictly ellipses. There is substantial flexibility in this method, as the designer can choose the number of coils, the regularity of the coil shapes, and any other input parameters to the coil design code. 
There is no particular reason the magnetic field will be quasisymmetric outside of the smaller high-aspect-ratio target volume, so this procedure tends to produce better quasisymmetry on axis than at the edge.
Since this method requires a coil design code, which takes at least $\sim$ 10 seconds to run in the case of REGCOIL,
as well as field line tracing to find the resulting surfaces, the computational cost is higher than that of the previous methods, though still very small compared to conventional stellarator optimization.
An example of this method will be shown at the end of section \ref{sec:QA}.


\section{Examples}
\label{sec:examples}


\subsection{Quasi-axisymmetry}
\label{sec:QA}

\begin{figure}
  \centering
    \includegraphics[width=2.8in,valign=t]{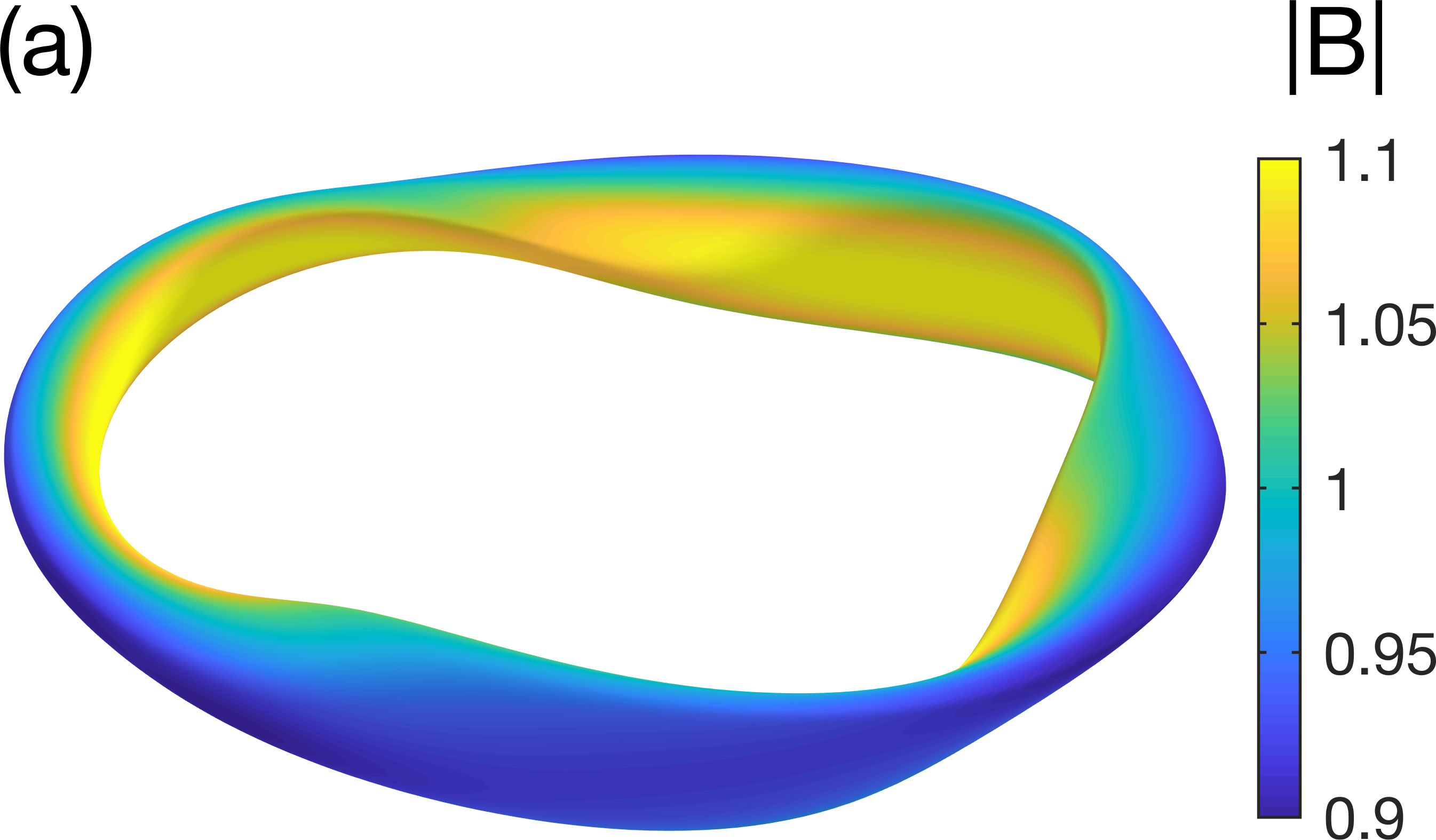}
  \includegraphics[width=2.2in,valign=t]{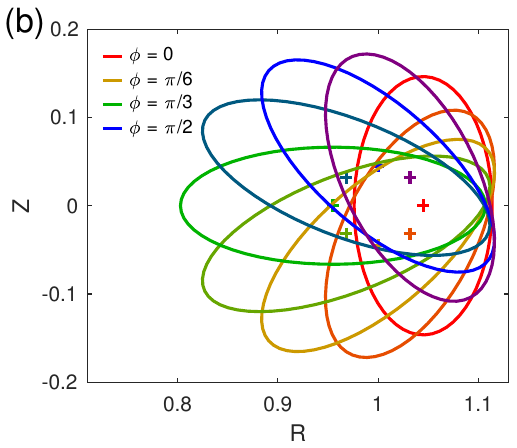}
  \caption{Quasi-axisymmetry example. (a) Flux surface shape computed by the procedure of sections \ref{sec:numerical} and \ref{sec:conversion}, taking aspect ratio = 10, showing $|B|$ computed by VMEC.
  (b) Cross sections of the flux surfaces at equally spaced values of $\phi$, with $+$ signs denoting the magnetic axis.}
\label{fig:QA1}
\end{figure}

\begin{figure}
  \centering
  \includegraphics[width=5.3in]{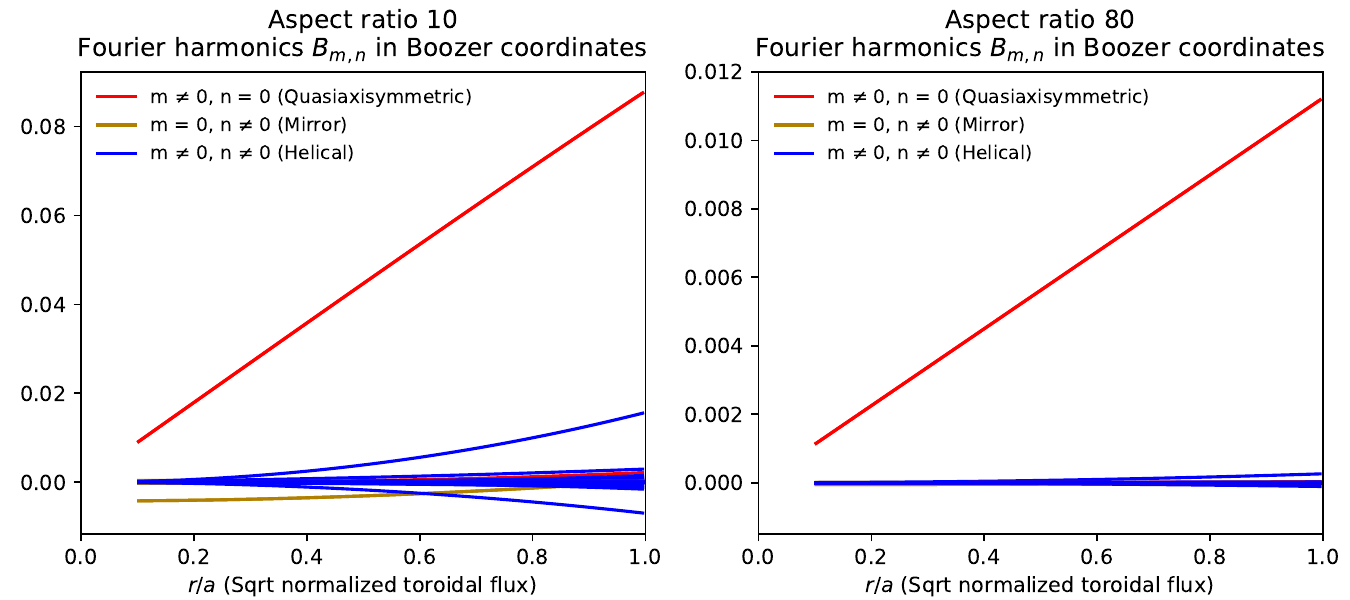}
  \caption{Fourier amplitudes $B_{m,n}(r)$ of the magnetic field magnitude $B(r,\theta,\boozertor)$ computed by BOOZ\_XFORM,
  for the  
  quasi-axisymmetric configuration of section \ref{sec:QA}.}
\label{fig:QA2}
\end{figure}

We now demonstrate the procedures of the previous sections
to construct a variety of quasisymmetric stellarator shapes.
We begin with an example of quasi-axisymmetry, considering the magnetic axis shape
\begin{equation}
\label{eq:axis_shape_QA}
R_0(\phi) = 1 + 0.045 \cos(3\phi), 
\hspace{0.5in}
z_0(\phi) = -0.045 \sin(3\phi),
\end{equation}
with $\bar\eta=-0.9$. 
We take $\sigma(0)=0$ (stellarator symmetry).
For this and the later examples, we consider a vacuum field, so $I_2=0$.
For these parameters, 
the numerical procedure above yields a rotational
transform 
$\iota=0.418$, 
and the maximum flux surface
elongation in the $R$-$z$ plane is found to be 
2.40.  
Hereafter we call $1/r$ the aspect ratio, since the average major radius is 1.
The flux surfaces for aspect ratio 10 are shown in figure \ref{fig:QA1}.
Supplying this surface as an input to VMEC, the resulting magnetic
field strength on the boundary is shown in figure \ref{fig:QA1}.a.
The Fourier spectrum of $B$ in Boozer coordinates at each flux surface is then computed using
the BOOZ\_XFORM code \citep{Sanchez}. The resulting spectra for aspect ratios 10 and 80
are shown in figure \ref{fig:QA2}. At aspect ratio 10,
the 
$(m,n)=(1,0)$
harmonic is dominant across all surfaces, as desired,
and the quality of the quasisymmetry increases as the aspect
ratio is increased. 
\changed{
For both aspect ratios shown, the largest symmetry-breaking mode at the edge is the mode $(m,n)=(2,-3)$.
Since modes of $B$ with given poloidal mode number $m$ have amplitude $\propto r^m$ near the axis \citep{GB1},
the symmetry-breaking on axis is dominated by modes with $m=0$ (shown in brown in figure \ref{fig:QA2}).}

\begin{figure}
  \centering
  \includegraphics[width=3in]{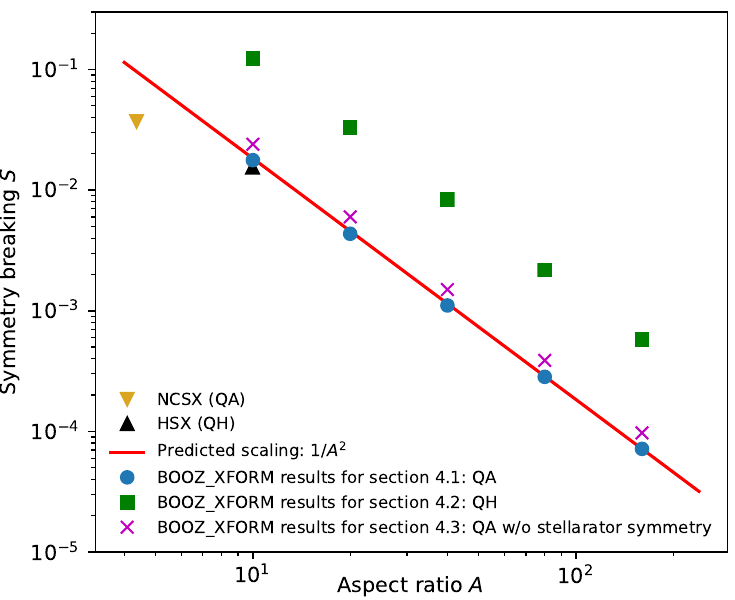}
  \caption{For all three examples presented in section \ref{sec:examples}, the
  symmetry-breaking Fourier components scale as $A^{-2}$ as predicted by theory.}
\label{fig:scaling}
\end{figure}

\begin{figure}
  \centering
  \includegraphics[width=3in]{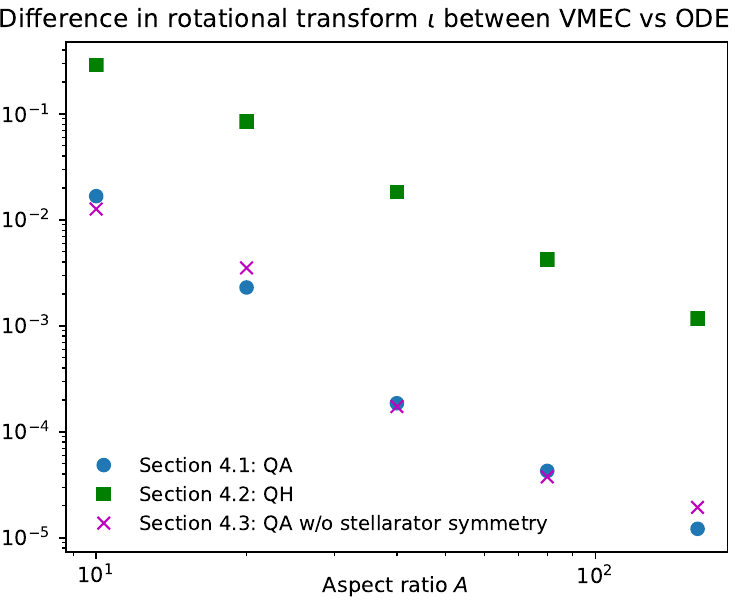}
  \caption{For all three examples presented in section \ref{sec:examples}, the
  rotational transform computed by VMEC converges to the value predicted by eq (\ref{eq:QA}) as the aspect ratio increases.}
\label{fig:iota}
\end{figure}

The theory here generates flux surface shapes
that give quasisymmetry to first order in the distance from the magnetic axis, and at next order in this distance there will be breaking of the symmetry. Therefore, the symmetry-breaking Fourier harmonics should scale as $1/A^2$ where $A$ is the aspect ratio. This scaling is verified in figure \ref{fig:scaling}. In this figure the amount of symmetry-breaking is measured by the quantity
\begin{equation}
S = \frac{1}{B_{0,0}}
\sqrt{\sum_{n/m\ne N/M}B_{m,n}^2}.
\end{equation}
As expected, the symmetric modes $B_{m,n}$ are found to scale as $1/A$ (not shown).
Similarly, figure \ref{fig:iota} shows that the rotational transform computed by VMEC converges to the value predicted by (\ref{eq:QA}) as the aspect ratio increases. For $A \ge 160$, the agreement extends to at least 5 digits.

\changed{Figure \ref{fig:scaling} includes a point for the quasi-axisymmetric design NCSX, which was obtained using conventional optimization. The NCSX point falls below the trend line, so it evidently has a somewhat better quality of quasisymmetry for its aspect ratio than the configurations constructed here.}

\begin{figure}
  \centering
    \includegraphics[width=2.6in,valign=m]{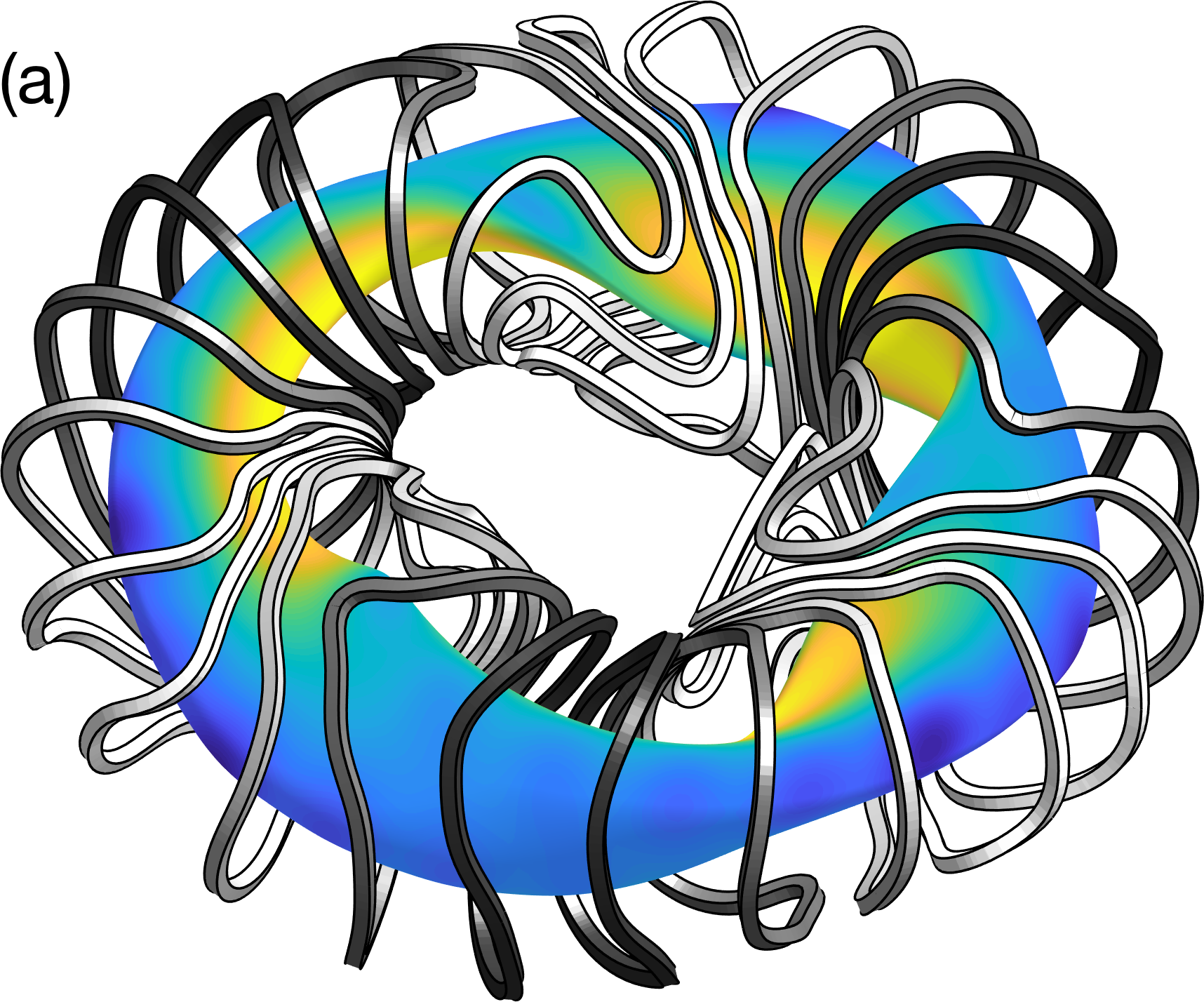}
    \includegraphics[width=2.6in,valign=m]{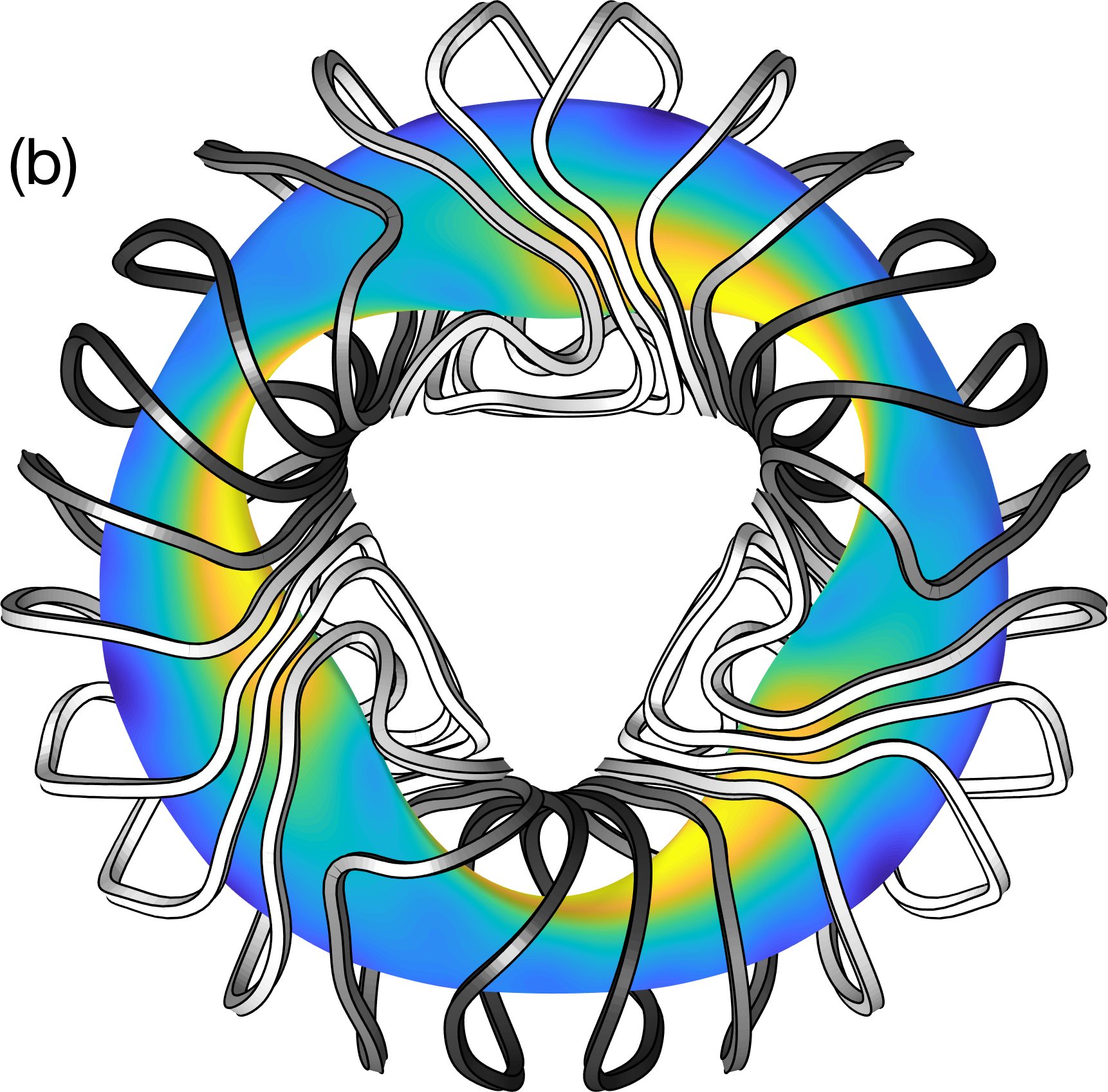}
  \includegraphics[width=5.3in]{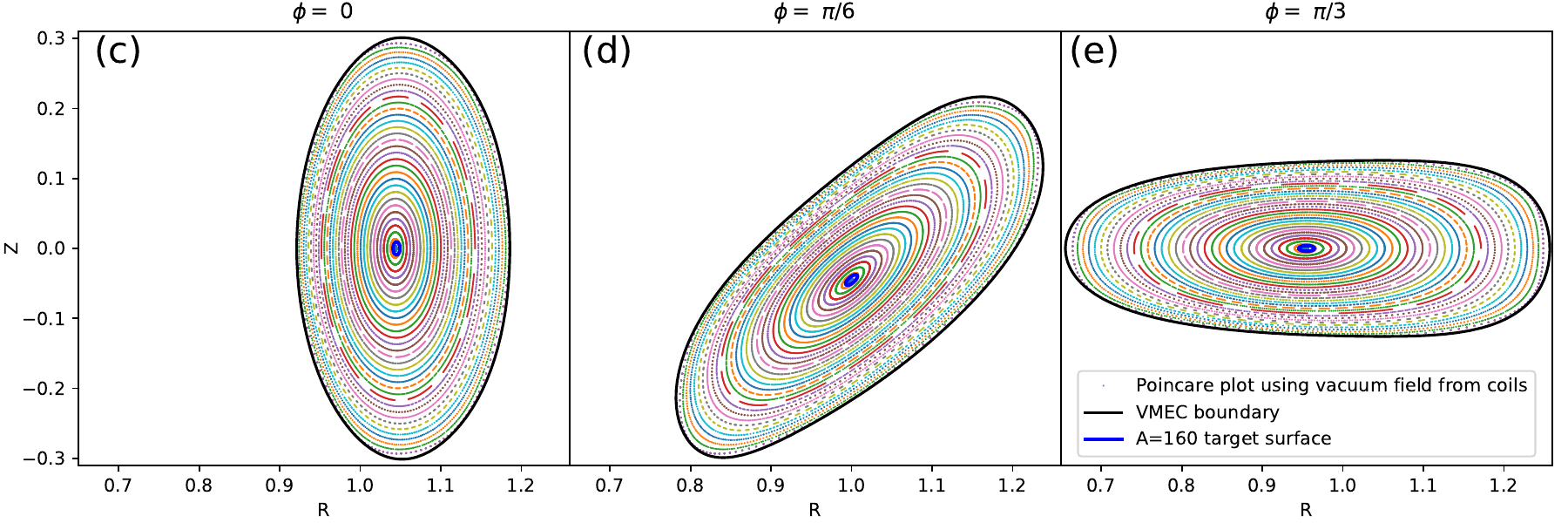}
    \includegraphics[width=2.7in,valign=m]{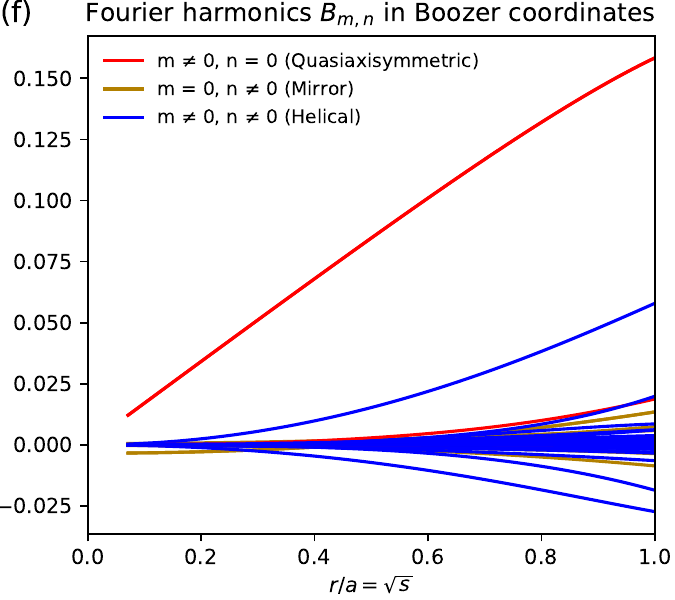}
  \includegraphics[width=2.5in,valign=m]{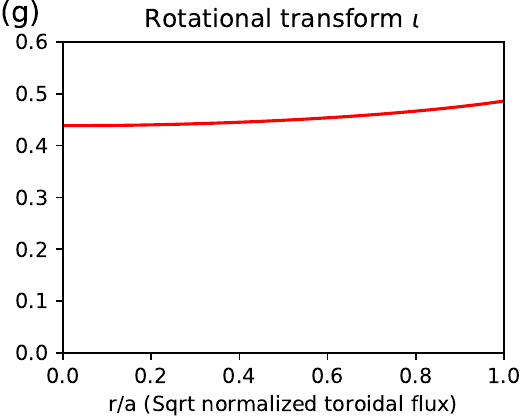}
  \caption{The aspect ratio 5 quasi-axisymmetic stellarator constructed by the procedure of section \ref{sec:conversion4}, using no optimization (aside from the REGCOIL linear least-squares problem). (a)-(b) Color indicates $B$ on the outermost flux surface, and the four unique coil shapes are shown with four shades of gray. (c)-(e) Poincare plots computed from the vacuum field of the coils, demonstrating good flux surfaces out to aspect ratio 5, at three toroidal angles. 
  (f) Boozer spectrum, demonstrating the quasi-axisymmetric mode is dominant.
  (g) Profile of $\iota$. }
\label{fig:QA_extrapolation}
\end{figure}

For a different approach to constructing a finite-aspect-ratio geometry
from the high-aspect-ratio theory, an example of the outward extrapolation method
of section \ref{sec:conversion4} is shown in figure
\ref{fig:QA_extrapolation},
again using the input axis shape (\ref{eq:axis_shape_QA}). First, an aspect ratio 160 shape is generated by the method of section \ref{sec:conversion}. Coil shapes
to produce this magnetic surface shape are then calculated
using the REGCOIL method \citep{regcoil}. For this method,
a coil winding surface is chosen by taking an aspect ratio 5 surface constructed using the method of section \ref{sec:conversion}, and expanding uniformly outward by one quarter of the average major radius. REGCOIL's
regularization parameter is chosen to be the smallest value for which there are no saddle coils, i.e. there are no local maxima or minima in the current potential. Next, 24 coil shapes (4 unique shapes, each repeated 6 times) are identified from uniformly spaced contours of the current
potential. A Poincare plot of the vacuum field produced by
these coils (figure \ref{fig:QA_extrapolation}.c-e) shows that good
flux surfaces exist out to an aspect ratio of 5.0 (using VMEC's definition of the major and minor radius).
The Fourier amplitudes of $B$ in Boozer coordinates are shown in figure \ref{fig:QA_extrapolation}.f,
showing the quasi-axisymmetric term is dominant, as desired. 
\changed{Again the largest symmetry-breaking mode at the edge is the mode $(m,n)=(2,-3)$.}
The symmetry-breaking harmonics reach a rather sizeable amplitude at the last closed flux surface, and 
no effort has been made to 
\changed{achieve other desirable physics properties such as a high MHD $\beta$ limit.}
However, this configuration required very little computational effort to compute,  compared to the hundreds or thousands of VMEC computations required for conventional optimization, and it could serve as a useful initial condition for conventional optimization.

\subsection{Quasi-helical symmetry}
\label{sec:QH}

For an example of quasi-helical symmetry, we consider the magnetic axis shape
\begin{equation}
R_0(\phi) = 1 + 0.265 \cos(4\phi), 
\hspace{0.5in}
z_0(\phi) = -0.21 \sin(4\phi).
\end{equation}
For this curve, the normal vector rotates poloidally in each field period, so solutions have quasi-helical symmetry rather than quasi-axisymmetry.
We also choose $\bar\eta=-2.25$ and $\sigma(0)=0$. For these parameters, 
the numerical procedure of sections \ref{sec:numerical}-\ref{sec:conversion} yields a rotational
transform $\iota=1.93$, and the maximum flux surface
elongation in the $R$-$z$ plane is found to be 2.52. 

\begin{figure}
  \centering
  \includegraphics[width=2.8in,valign=t]{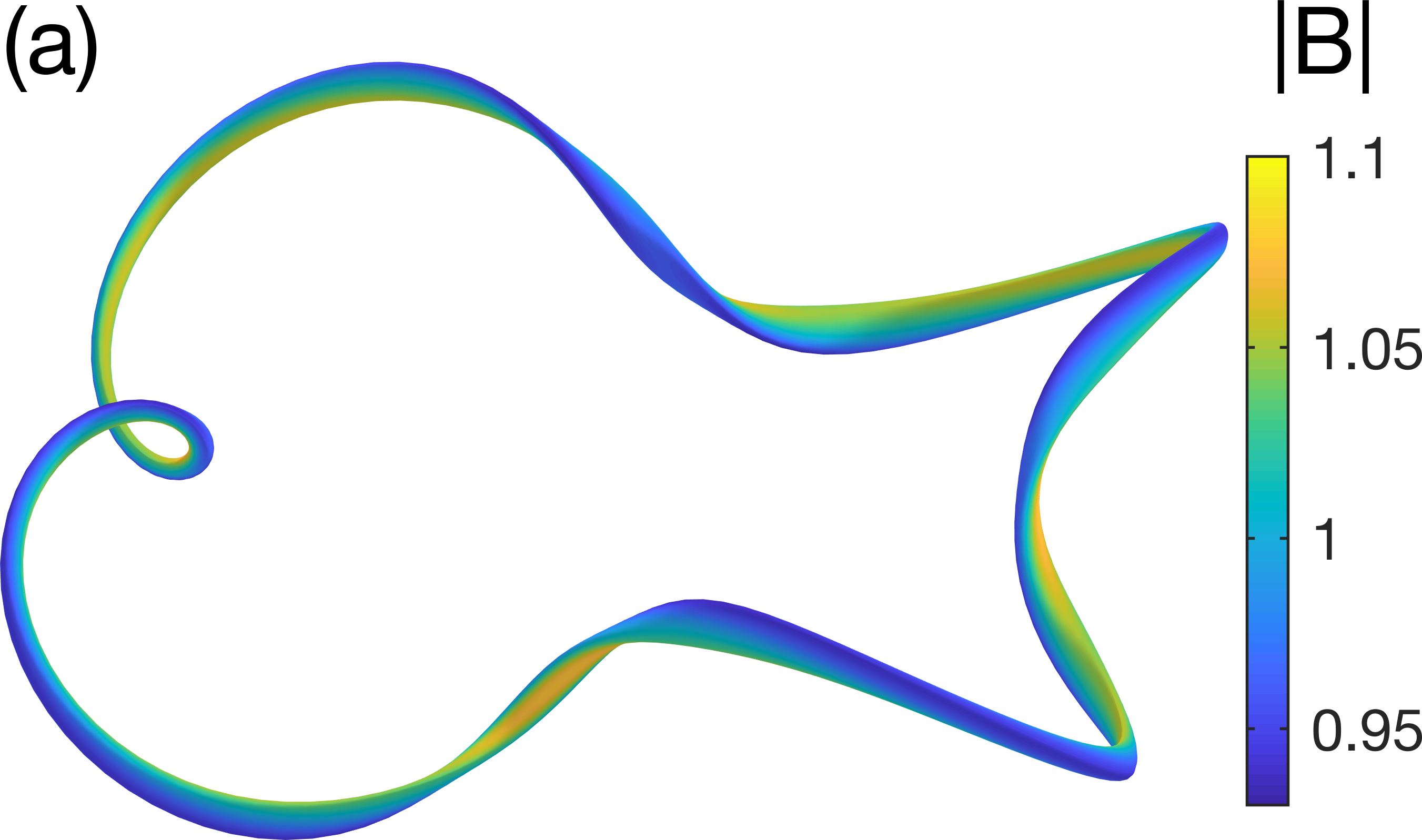}
  \includegraphics[width=2.2in,valign=t]{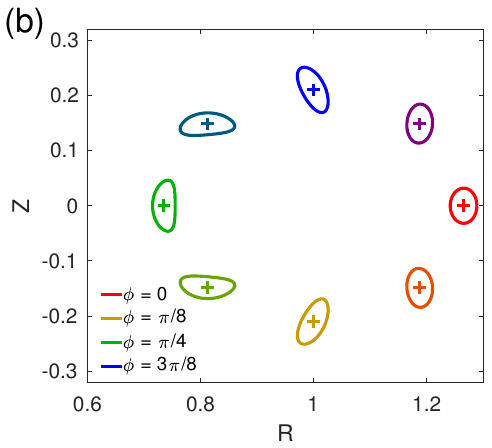}
  \caption{Quasi-helical symmetry example. (a) Flux surface shape computed by the procedure of sections \ref{sec:numerical} and \ref{sec:conversion2}, taking aspect ratio = 40, showing $|B|$ computed by VMEC.
  (b) Cross sections of the flux surfaces at equally spaced values of $\phi$, with $+$ signs denoting the magnetic axis.}
\label{fig:QH1}
\end{figure}

\begin{figure}
  \centering
\includegraphics[width=5.3in]{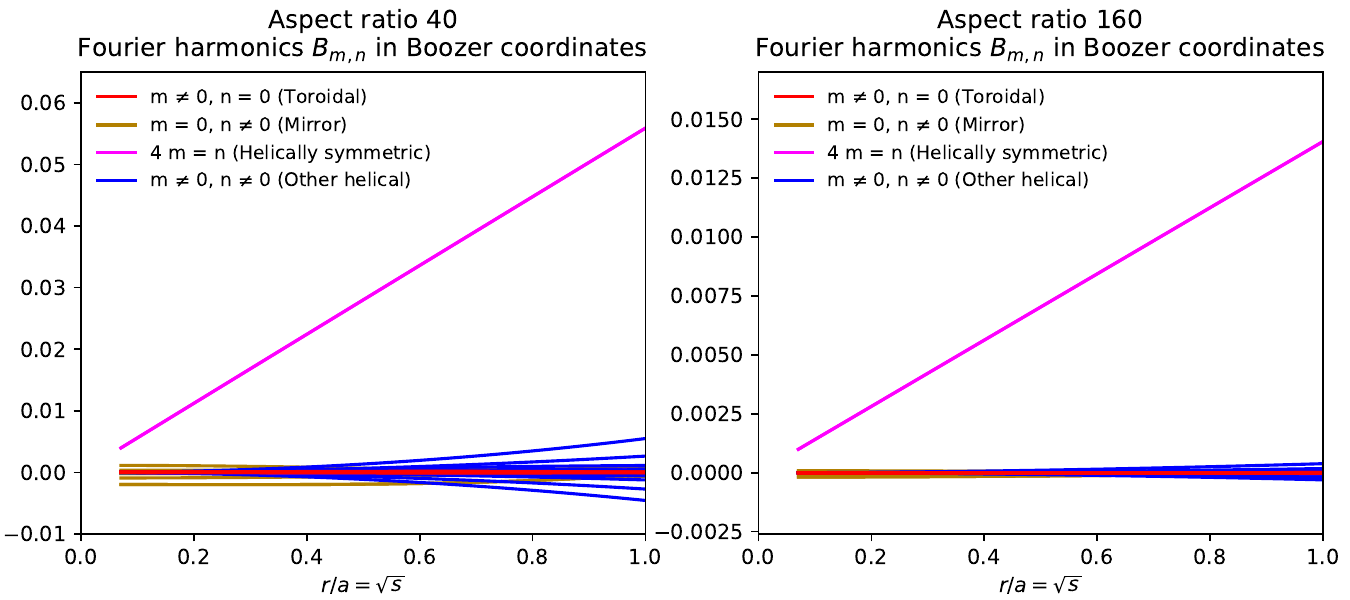}
  \caption{Fourier amplitudes $B_{m,n}(r)$ of the magnetic field magnitude $B(r,\theta,\boozertor)$ computed by BOOZ\_XFORM,
  for the  
  quasi-helically symmetric configuration of section \ref{sec:QH}.}
\label{fig:QH2}
\end{figure}

The flux surfaces for aspect ratio 40, computed using the method of section \ref{sec:conversion2}, are shown in figure \ref{fig:QH1}.
Due to the strongly shaped axis in this example, the flux surface cross-sections
in the $R$-$z$ plane become visibly different from ellipses even at this high aspect ratio.
Note that the cross-sections in the plane perpendicular to the magnetic axis are perfectly elliptical, and the cross-sections in the $R$-$z$ plane approach ellipses as the aspect ratio is raised.
The spectra of $B$ in Boozer coordinates for aspect ratios 40 and 160
are shown in figure \ref{fig:QH2}. 
\changed{The largest symmetry-breaking mode at the edge is the mode $(m,n)=(2,-16)$.}

As pointed out by \cite{GB1}, the relevant ratio for breaking of quasisymmetry is the minor radius
divided by the scale length of the magnetic axis's Frenet-Serret frame (e.g. $1/\kappa$, $1/\tau$), not the conventional aspect ratio. For axis shapes consistent with quasi-helical symmetry, where the normal vector rotates about the axis, the scale lengths of the axis Frenet-Serret frame are smaller than for axes consistent with quasi-axisymmetry at comparable major radius. Therefore, quasi-helical symmetry is limited to higher conventional aspect ratios than quasi-axisymmetry. This trend is apparent in a comparison of the examples of sections \ref{sec:QA}-\ref{sec:QH}. The peak axis curvature and torsion are roughly twice as large in the latter compared to the former, and this ratio is squared in the symmetry breaking. Indeed, figure \ref{fig:QA2}.a for quasi-axisymmetry at aspect ratio 10 has comparable symmetry breaking to figure \ref{fig:QH2}.a for quasi-helical symmetry at aspect ratio 40.

\changed{Figure \ref{fig:scaling} includes a point for the quasi-helically symmetric experiment HSX, which was designed using conventional optimization. (Coil ripple is not included for the HSX and NCSX configurations in the figure; the values of $S$ for HSX and NCSX are nearly unchanged on the scale of the figure if coil ripple is included.) HSX has symmetry breaking that is an order of magnitude smaller than the configuration generated here at comparable aspect ratio.
The fact that conventional optimization results in lower symmetry breaking than the construction here is not surprising,
given that the construction is limited to producing shapes with elliptical cross section.}


\subsection{Case without stellarator symmetry}
\label{sec:asymmetric}

\begin{figure}
  \centering
  \includegraphics[width=2.8in,valign=t]{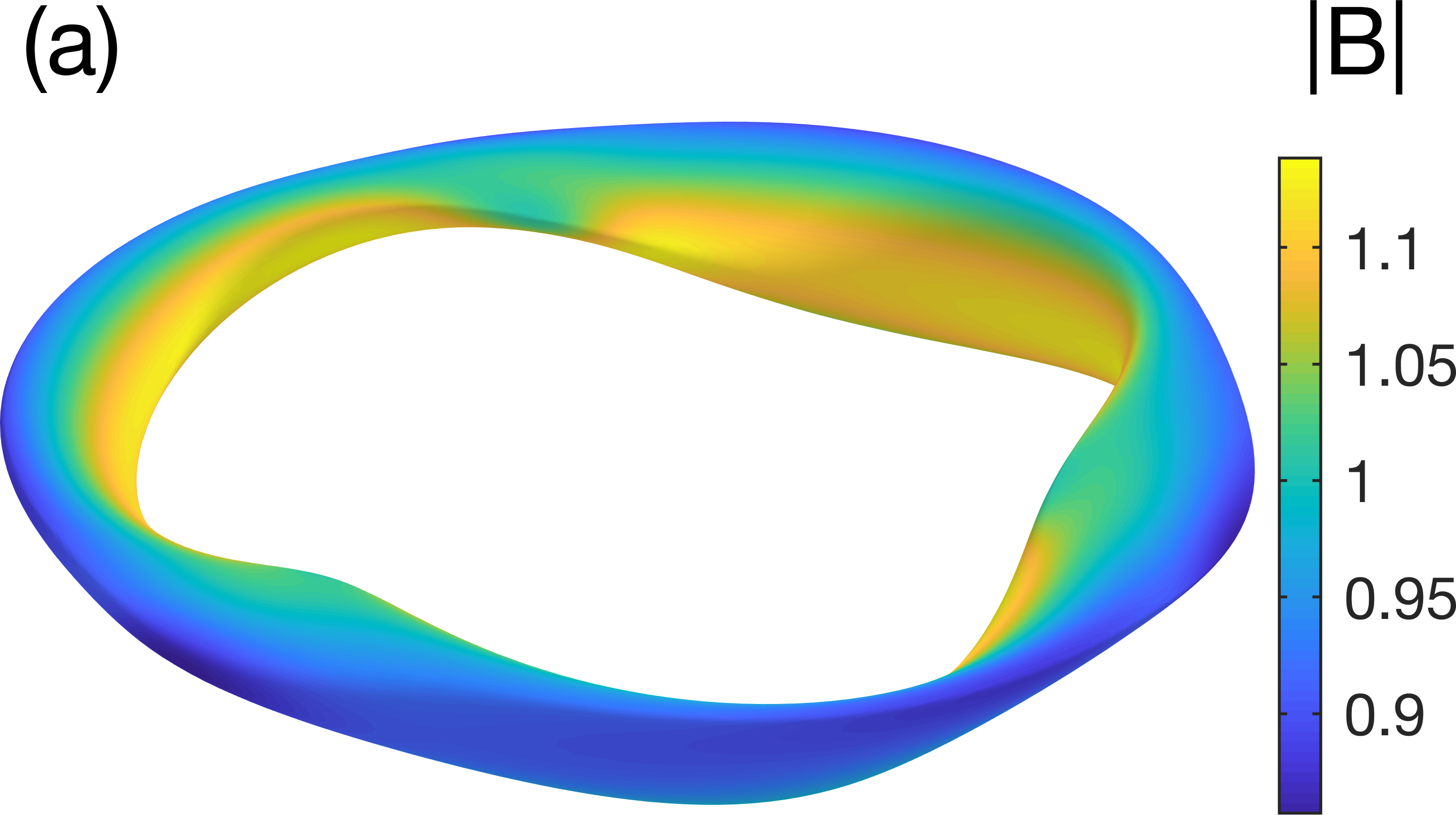}
  \includegraphics[width=2.2in,valign=t]{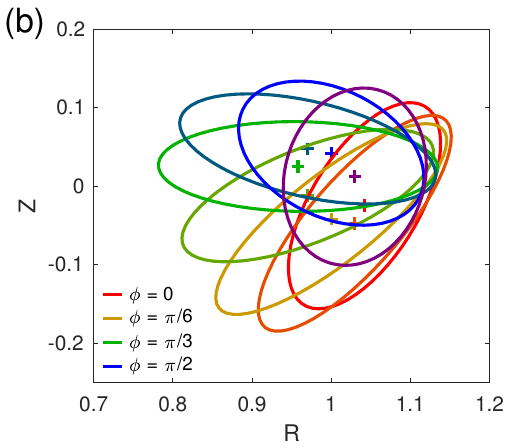}
  \caption{Quasi-axisymmetric stellarator without stellarator symmetry. (a) Flux surface shape computed by the procedure of sections \ref{sec:numerical} and \ref{sec:conversion}, taking aspect ratio = 10, showing $|B|$ computed by VMEC.
  (b) Cross sections of the flux surfaces at equally spaced values of $\phi$, with $+$ signs denoting the magnetic axis.}
\label{fig:asymmetric1}
\end{figure}

\begin{figure}
  \centering
  \includegraphics[width=5.3in]{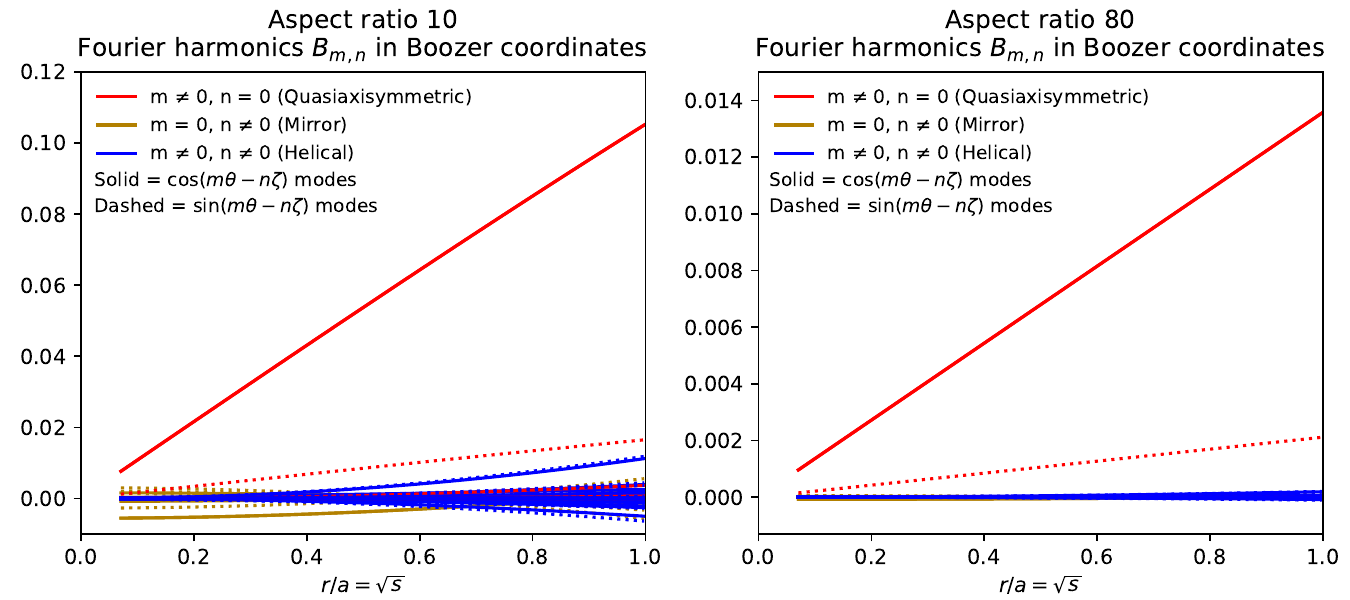}
  \caption{Fourier amplitudes $B_{m,n}(r)$ of the magnetic field magnitude $B(r,\theta,\boozertor)$ computed by BOOZ\_XFORM,
  for the non-stellarator-symmetric 
  quasi-axisymmetric configuration of section \ref{sec:asymmetric}.}
\label{fig:asymmetric2}
\end{figure}

There is no reason a stellarator with quasisymmetry
must also possess stellarator symmetry.
For instance, a tokamak with a single null is quasiaxisymmetric but not stellarator symmetric.
Plasma shapes that lack stellarator symmetry are of interest
since the turbulent momentum flux is predicted to be larger by a factor $\sim 1/\rho_*$ than in stellarator symmetric shapes, meaning the intrinsic rotation is larger \citep{Peeters, Parra, Sugama}. The resulting rotation and/or rotation shear may improve plasma stability. 
\changed{While quasisymmetry reduces the strong damping of flows otherwise typical of stellarators, significant flow still requires a drive, and turbulent momentum transport associated with broken stellarator symmetry could provide such a drive.}
In the model considered here, stellarator symmetry
can be broken by specifying a non-stellarator-symmetric axis shape, or by specifying a nonzero $\sigma(0)$, or both.
Here we present an example with both sources of symmetry-breaking. We take the magnetic axis shape to be
\begin{equation}
R_0(\phi) = 1 + 0.042 \cos(3\phi), 
\hspace{0.5in}
z_0(\phi) = -0.042 \sin(3\phi) - 0.025 \cos(3\phi),
\end{equation}
with 
$\bar\eta=-1.1$ and $\sigma(0)=-0.6$. 
For these parameters, 
the numerical procedure above yields a rotational
transform 
$\iota=0.311$,
and the maximum flux surface
elongation in the $R$-$z$ plane is found to be 
3.29.
The flux surface shape for $A=10$ is displayed in figure \ref{fig:asymmetric1}, and the Boozer spectra
for $A=10$ and $A=80$ are shown in figure \ref{fig:asymmetric2}. In figure \ref{fig:asymmetric2}, it can be seen that $B$ has a significant $\sin\theta$ component (red dotted line) which is not stellarator-symmetric but which preserves quasi-axisymmetry. 
\changed{As with the stellarator-symmetric quasi-axisymmetric example, the largest symmetry-breaking mode at the edge is the mode $(m,n)=(2,-3)$.}
The $1/A^2$ scaling of the quasisymmetry-breaking harmonics is again plotted in figure \ref{fig:scaling}, and the convergence of the VMEC rotational transform to the predicted value as $A\to\infty$ is shown in figure \ref{fig:iota}. Generally the properties of this family of configurations are quite similar to the stellarator-symmetric and quasi-axisymmetric configurations of section \ref{sec:QA}.

\section{Discussion and Conclusions}

While quasisymmetric stellarator shapes have been found previously by applying
black-box optimization methods to minimize the departure from quasisymmetry,
such methods are computationally demanding and do not provide comprehensive
information about the landscape of all possible solutions. Here we have
demonstrated a complimentary approach in which quasisymmetric stellarator shapes
can be parameterized and computed extremely rapidly (< 1 ms on a laptop), enabling exhaustive
high-resolution parameter scans and insight into the size of the solution space.
We have demonstrated that this approach is a practical method to generate
configurations that can be examined numerically with VMEC and other physics codes.
As part of this demonstration, we have shown using BOOZ\_XFORM that the Boozer-coordinate Fourier spectra
of the resulting equilibria are indeed dominated by a single helicity. We have further demonstrated that the symmetry-breaking harmonics scale as $r^2$ as expected.

Although the ``optimization-free'' approach here requires solving a nonlinear equation (\ref{eq:QA}), the numerical solution
is extremely robust since the problem can be formulated so a unique solution is guaranteed to exist.
As proved in the appendix, \emph{every} magnetic axis shape with nonvanishing curvature admits an infinity of (first-order) quasisymmetric flux surface shapes surrounding it, each labeled by the three numbers $\bar\eta$, $I_2$, and $\sigma(0)$. Given any values for these three numbers, as long as $\bar\eta\ne 0$, and given any axis shape for which the curvature does not vanish, there is exactly one first-order quasisymmetric shape (as a function of $r$.)
However, much of this solution space is not interesting since the elongation of the surfaces 
is impractically high. The space of stellarator-symmetric solutions is significantly smaller than the space of all solutions both
because the space of axis shapes is restricted and also since $\sigma(0)$ must be 0.

While the calculations here are limited to high aspect ratio, any
stellarator with a low aspect ratio boundary will have a region close to the magnetic 
axis in which the local aspect ratio is high. Hence the results
here describe the core of any quasisymmetric stellarator, even those with low aspect ratio at the plasma boundary.
The observed accuracy of the solutions constructed here is consistent with the popular
wisdom that good quasi-axisymmetry can be achieved at much lower aspect ratio
than quasi-helical symmetry.
It is likely that by extending the method here to second order in $r$,
using equations in the appendix of \cite{GB2}, the accuracy of these
parameterized solutions could be extended to lower aspect ratio.



We are grateful to 
Harold Weitzner for pointing out the possibility of singularity in (\ref{eq:PQ}).
The idea of a non-stellarator-symmetric quasisymmetric stellarator was suggested by Greg Hammett.
This work was supported by the
U.S. Department of Energy, Office of Science, Office of Fusion Energy Science,
under award numbers DE-FG02-93ER54197 and DE-FG02-86ER53223.
This work was also supported by a grant from the Simons Foundation (560651, ML).


\appendix

\section{Existence and uniqueness of solutions to the ODE}

The problem for first-order quasisymmetry (\ref{eq:QA}) can be stated as
\begin{equation}
\frac{d\sigma}{d\boozertor}+\iota(P + \sigma^2) + Q = 0,
\hspace{0.5in}
\sigma(0) \;\;\mbox{given},
\label{eq:PQ}
\end{equation}
where $\sigma(\boozertor)$, $P(\boozertor)$, and $Q(\boozertor)$ are $2\pi$-periodic functions, $P>0$,
and $\iota$ and $\sigma(0)$ are constants. 
(Without loss of generality, the shift $-N$ to $\iota$ is dropped in this appendix to simplify notation.)
Here we prove that for given $P$, $Q$, and $\sigma(0)$,
assuming that $P$ and $Q$ are integrable and bounded,
a periodic solution $\{\iota,\; \sigma(\boozertor)\}$ to (\ref{eq:PQ}) exists and it is unique.

Note that if the problem is posed instead with $\iota$ as given and $\sigma(0)$ as part of the solution, rather than the other way around, then there may be zero, one, or two solutions. In this alternative formulation there can never be more than two solutions (\cite{Pliss}, page 102).


\subsection{Uniqueness}
\label{sec:uniqueness}

\begin{figure}
  \centering
  \includegraphics[width=2.5in,valign=t]{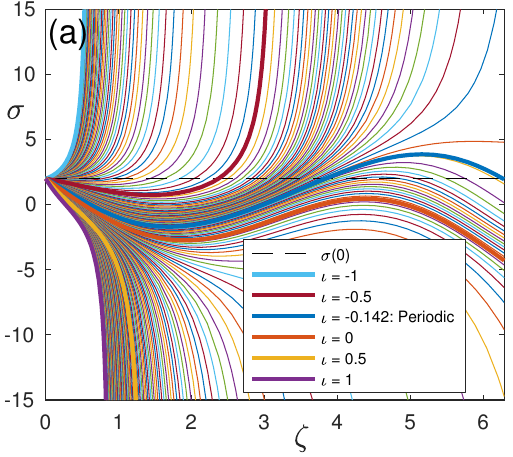}
  \includegraphics[width=2.5in,valign=t]{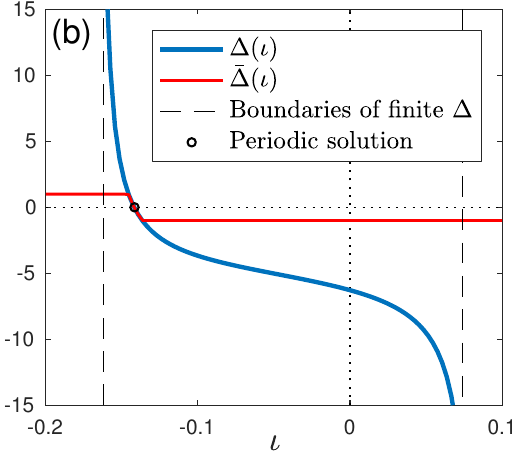}
  \caption{(a) Solutions of the ODE (\ref{eq:PQ}),
  interpreted as an initial value problem, for
  $P=2+\sin(2\boozertor)$, $Q=1+3\cos\boozertor$, $\sigma(0)=2$, and various $\iota\in[-1,1]$. (b) Demonstration that $\Delta(\iota)=\sigma(2\pi)-\sigma(0)$ is a
  monotonically decreasing function of $\iota$, and illustration of the function $\bar\Delta(\iota)$ of section \ref{sec:continuity2}, for the same parameters as (a).}
\label{fig:ODE}
\end{figure}

Returning to the original formulation of (\ref{eq:PQ}) with $\sigma(0)$ as an input and $\iota$ as an output,  we will first prove that no more than one solution can exist. 
For the moment, we relax the requirement that $\sigma(\boozertor)$ be periodic, so that for any given $\iota$,
(\ref{eq:PQ}) becomes an initial value problem, which has a unique, finite, and generally non-periodic solution
$\sigma(\boozertor)$ in some neighborhood of $\boozertor=0$. 
Solutions for a particular choice of $P$, $Q$, and $\sigma(0)$ are
shown in figure \ref{fig:ODE}.a.
Note that the solution of this initial value problem 
may not extend all the way to $\boozertor=2\pi$ since 
it may diverge to $\pm \infty$ beforehand, 
as can be seen from the analytic solution
in the case of constant $P$ and $Q$:
\begin{align}
\sigma(\boozertor) = -\sqrt{(Q+P\iota)/\iota}\tan\left( \boozertor\sqrt{(Q+P\iota)\iota}
-\tan^{-1}\left(\sigma(0)\sqrt{\iota/(Q+P\iota)}\right)\right).
\end{align}
Returning to the case of general $P>0$ and $Q$, suppose $\sigma_0(\boozertor)$ is the solution to the initial value problem for $\iota=\iota_0$,
and $\sigma_1(\boozertor)$ is the solution for $\iota=\iota_1$.
Subtracting (\ref{eq:PQ}) for these two solutions,
\begin{align}
0&=
\frac{d(\sigma_1-\sigma_0)}{d\boozertor} + (\iota_1-\iota_0)P + \iota_1\sigma_1^2 - \iota_0\sigma_0^2
\nonumber \\
&=
\frac{d(\sigma_1-\sigma_0)}{d\boozertor} + (\iota_1-\iota_0)(P + \sigma_0^2 )
+\iota_1(\sigma_1- \sigma_0)(\sigma_1+\sigma_0).
\label{eq:difference}
\end{align}
This equation may be integrated using an integrating factor to give
\begin{align}
\sigma_1(\boozertor)-\sigma_0(\boozertor) 
=-(\iota_1 - \iota_0) F(\boozertor),
\label{eq:monotonic}
\end{align}
where
\begin{align}
F(\boozertor)=&
\exp\left( -\iota_1 \int_0^\boozertor d\boozertor' \left[ \sigma_1(\boozertor')+\sigma_0(\boozertor')\right]\right) \nonumber \\
&\times \int_0^\boozertor d\boozertor'' \left[ P(\boozertor'')+\sigma_0^2(\boozertor'')\right]
\exp\left( \iota_1 \int_0^{\boozertor''}d\boozertor' \left[ \sigma_1(\boozertor')+\sigma_0(\boozertor')\right]\right).
\end{align}
Using $P>0$ it can be seen that $F(\boozertor)>0$ for any positive $\boozertor$ for which the initial value solutions exist, so $\iota_1 > \iota_0$ implies $\sigma_1(\boozertor) < \sigma_0(\boozertor)$.
That is, $\sigma(\boozertor)$ is a strictly monotonically decreasing function of $\iota$ at any $\boozertor$ for which the initial value solution exists.
This is true in particular at $\boozertor=2\pi$. Defining $\Delta(\iota) = \sigma(2\pi) - \sigma(0)$ (defined at any $\iota$ for which the initial value solutions do extend to $2\pi$), we then have
\begin{align}
\Delta(\iota_1) - \Delta(\iota_0) = -(\iota_1 - \iota_0) F(2\pi).
\end{align}
Using $F>0$, $\Delta(\iota)$ is a strictly monotonically decreasing function.
Figure \ref{fig:ODE}.b shows $\Delta(\iota)$ for the particular parameters
of figure \ref{fig:ODE}.a, and this monotonicity is apparent.
Thus, no more than a single value of $\iota$ can exist for which $\Delta(\iota)=0$,
corresponding to a periodic $\sigma(\boozertor)$.

\subsection{Bounded solutions}
\label{sec:bounded}


To prove that at least one solution to (\ref{eq:PQ}) exists,
let us prove several intermediate results that will be needed, beginning
with the following proposition. Suppose when $\iota = \iota_0$,
the initial value problem (\ref{eq:PQ}) with some
given initial condition $\sigma(0)$ has a bounded solution $\sigma_0(\boozertor)$ throughout $\boozertor\in[0,2\pi]$. Then there exists some $d>0$
such that for all $\iota$ satisfying $|\iota - \iota_0| < d$,
then $\sigma$ solving the initial value problem (\ref{eq:PQ}) with $\iota$ and the same initial condition remains bounded throughout $\boozertor\in[0,2\pi]$.
In other words, for any $\iota_0$ that yields a solution that is non-singular,
there are nearby values of $\iota$ that also avoid singularity.
Put yet another way, if $\mathcal{B}$ is the set of values of $\iota$ that 
yield bounded solutions to the initial value problem (for a given $\sigma(0)$), then $\mathcal{B}$ is open.

To prove this proposition, it is useful to consider the pair of solutions $\{\iota_0, \sigma_0\}$ and $\{\iota, \sigma\}$ which are both finite up to some $\boozertor$, and write
(\ref{eq:monotonic}) as
\begin{align}
\label{eq:monotonicAlternate}
&\left[ \sigma(\boozertor) - \sigma_0(\boozertor)\right]
\exp\left(
\iota \int_0^\boozertor d\boozertor' \left[ \sigma(\boozertor') - \sigma_0(\boozertor')\right]\right) \\
&=(\iota_0 - \iota)
\exp\left( -2\iota \int_0^\boozertor d\boozertor' \sigma_0(\boozertor')\right)
\int_0^\boozertor d\boozertor'' \left[ P(\boozertor'')+\sigma_0^2(\boozertor'')\right]
\exp\left( \iota \int_0^{\boozertor''}d\boozertor' \left[ \sigma(\boozertor')+\sigma_0(\boozertor')\right]\right).\nonumber
\end{align}
Recognizing the left hand side as a total derivative
$\iota^{-1} (d/d\boozertor) \exp(\ldots)$, and integrating,
\begin{align}
\label{eq:trickyBound}
&\exp\left(
\iota \int_0^\boozertor d\boozertor' \left[ \sigma(\boozertor') - \sigma_0(\boozertor')\right]\right)
=1+\iota (\iota_0 - \iota)
\int_0^{\boozertor} d\boozertor'''
\exp\left( -2\iota \int_0^{\boozertor'''} d\boozertor' \sigma_0(\boozertor')\right)
\hspace{0in}\\
&\hspace{1.5in}\times \int_0^{\boozertor'''} d\boozertor'' \left[ P(\boozertor'')+\sigma_0^2(\boozertor'')\right]
\exp\left( \iota \int_0^{\boozertor''}d\boozertor' \left[ \sigma(\boozertor')+\sigma_0(\boozertor')\right]\right).\nonumber
\end{align}

We now consider three cases, depending on the sign of $\iota_0$,
beginning with the case $\iota_0<0$. If $\iota$ lies in $(\iota_0,0)$,
then the fact that $\sigma$ is a monotonically decreasing function of $\iota$ at each $\boozertor$ means that $\sigma$ is bounded between $\sigma_0$ and
\begin{equation}
\sigma_Q(\boozertor) = \sigma(0) - \int_0^\boozertor d\boozertor' Q(\boozertor'),
\end{equation}
the solution of the initial value problem for $\iota=0$. As $\sigma_0$ and $\sigma_Q$ are bounded throughout $[0,2\pi]$, $\sigma$ cannot be unbounded. On the other hand,
if $\iota < \iota_0$, then $\sigma > \sigma_0$ and (\ref{eq:trickyBound}) imply
\begin{align}
\label{eq:Ybound}
&\exp\left(
\iota \int_0^\boozertor d\boozertor' \left[ \sigma(\boozertor') - \sigma_0(\boozertor')\right]\right)
> 1-Y(\boozertor)
\end{align}
where
\begin{align}
Y(\boozertor) =&\iota (\iota - \iota_0)
\int_0^{\boozertor} d\boozertor'''
\exp\left( -2\iota \int_0^{\boozertor'''} d\boozertor' \sigma_0(\boozertor')\right)
\hspace{0in}\\
&\hspace{1.0in}\times \int_0^{\boozertor'''} d\boozertor'' \left[ P(\boozertor'')+\sigma_0^2(\boozertor'')\right]
\exp\left( 2\iota \int_0^{\boozertor''}d\boozertor' \sigma_0(\boozertor')\right).\nonumber
\end{align}
Note $Y>0$. If $Y<1$, then the reciprocal of (\ref{eq:Ybound}) can be 
applied to (\ref{eq:monotonicAlternate}) to obtain
\begin{align}
\label{eq:trickyBound2}
\sigma(\boozertor) < &\sigma_0(\boozertor) - \frac{\iota-\iota_0}{1-Y}
\exp\left(-2\iota \int_0^\boozertor d\boozertor' \sigma_0(\boozertor')\right) \\
&\hspace{0.7in}\times\int_0^\boozertor d\boozertor'' \left[ P(\boozertor'')+\sigma_0^2(\boozertor'')\right]
\exp\left(2\iota \int_0^{\boozertor''}d\boozertor' \sigma_0(\boozertor')\right) .\nonumber
\end{align}
Therefore, $\sigma$ is bounded between $\sigma_0$ and the right hand side of (\ref{eq:trickyBound2}), both of which are finite as long as $Y$ is bounded away from 1. 
To bound $Y$, it is convenient to require
$|\iota - \iota_0| < 1$, so
\begin{align}
\iota_0 \sigma_0 - |\sigma_0| < \iota \sigma_0 
< \iota_0 \sigma_0 + |\sigma_0|,
\end{align}
and $1/(-\iota) >1/(1-\iota_0)$.
Then requiring $|\iota - \iota_0|<d_-$ where
\begin{align}
\label{eq:d}
d_- =& \frac{1}{2(1-\iota_0)}\min_\boozertor \left\{
\int_0^\boozertor d\boozertor'''
\exp\left(-2\int_0^{\boozertor'''} d\boozertor' \left[ \iota_0 \sigma_0(\boozertor') -
|\sigma_0(\boozertor')|\right]\right) \right. \\
&\hspace{0.9in} \left.\times
\int_0^{\boozertor'''} d\boozertor'' \left[ P(\boozertor'') + \sigma_0(\boozertor'')\right]
\exp \left( 2\int_0^{\boozertor''}d\boozertor' \left[ \iota_0\sigma_0(\boozertor') + |\sigma_0(\boozertor')|\right]\right) \right\}^{-1}, \nonumber
\end{align}
where $\min_\boozertor$ indicates a minimum over $\boozertor\in[0,2\pi]$,
it follows that $Y<1/2$. So in summary, whenever $\iota_0<0$,
if $\iota$ satisfies $|\iota - \iota_0| < d$ where $d = \min(1, -\iota_0, d_-)$, then $\sigma$ will be bounded between two functions that are nonsingular throughout $\boozertor\in[0,2\pi]$: $\sigma_0$ and either $\sigma_Q$ or the right hand side of (\ref{eq:trickyBound2}). Hence $\sigma$ cannot be unbounded.

The case $\iota_0>0$ can be analyzed just as the case $\iota_0<0$. 
This time the final bound obtained is $d = \min(1, \iota_0, d_+)$,
where $d_+$ is defined exactly as $d_-$ in (\ref{eq:d}) but with $1/(1-\iota_0) \to 1/(1+\iota_0)$. For the final case, $\iota_0=0$, then the upper bound (\ref{eq:trickyBound2}) applies, as does a lower bound analogous 
to (\ref{eq:trickyBound2}) from the $\iota_0>0$ case. Therefore a suitable bound on $\iota$ is 
$d = \min(1,d_-, d_+)$.


\subsection{Continuity of $\Delta(\iota)$}
\label{sec:continuity}

Before proceeding to prove that at least one solution to (\ref{eq:PQ}) exists, we need to prove that $\Delta(\iota)$
is continuous at every $\iota = \iota_0$ for which (\ref{eq:PQ}) can be integrated to $\boozertor=2\pi$. To do this, we continue to allow non-periodic $\sigma$, we let $\sigma_+(\boozertor)$ be the generally non-periodic solution of (\ref{eq:PQ}) with $\iota = \iota_0+d$, and we let $\sigma_-(\boozertor)$ be the solution with $\iota=\iota_0-d$.
Here, $d>0$ is the quantity defined in section \ref{sec:bounded},
guaranteeing $\sigma_-$ and $\sigma_+$ are finite throughout $\boozertor \in [0,2\pi]$. Now consider any $\iota_1$ in the interval $(\iota_0,\iota_0+d)$ with associated solution $\sigma_1$.
From section \ref{sec:bounded}, we know
$\sigma_1$ is finite throughout $\boozertor \in [0,2\pi]$.
From (\ref{eq:monotonic}) and $F(\boozertor)>0$ for $\boozertor>0$, then $\sigma_+ < \sigma_1 < \sigma_0$ for any $\boozertor>0$.
Defining
\begin{equation}
A(\boozertor) = -(|\iota_0|+d) \left[|\sigma_0(\boozertor)| + \max\left(|\sigma_0(\boozertor)|, \; |\sigma_-(\boozertor)|, \; |\sigma_+(\boozertor)|\right)\right],
\end{equation}
and noting that $a<b<c$ implies $|b|<\max(|a|,\,|c|)$ for any numbers $(a,b,c)$,
it can be seen that $\iota_1(\sigma_1+\sigma_0) > A$. Therefore (\ref{eq:difference}) implies
\begin{align}
\frac{d(\sigma_1-\sigma_0)}{d\boozertor} 
>  - (\iota_1-\iota_0)(P + \sigma_0^2 )
-(\sigma_1- \sigma_0)A.
\label{eq:A1}
\end{align}
Using an integrating factor as before and integrating over $[0,2\pi]$,
\begin{align}
\Delta(\iota_1)-\Delta(\iota_0) > - (\iota_1-\iota_0) C
\label{eq:continuity_bound}
\end{align}
where
\begin{align}
C &= \exp\left( -\int_0^{2\pi} d\boozertor A(\boozertor)\right)
\int_0^{2\pi} d\boozertor \left[ P(\boozertor) + \sigma_0(\boozertor)^2 \right]
\exp\left( \int_0^{\boozertor}d\boozertor' A(\boozertor')\right).
\end{align}
Similarly, if $\iota_1$ is in the interval $(\iota_0-d,\iota_0)$, then $\sigma_0 < \sigma_1 < \sigma_-$ for any $\boozertor>0$.
Again $\iota_1(\sigma_1+\sigma_0) > A$, and so (\ref{eq:difference}) implies
(\ref{eq:A1}) with the direction of inequality reversed. Then 
(\ref{eq:continuity_bound}) follows with the direction of inequality reversed.
Therefore, given any $\epsilon>0$, we can define
$\delta(\epsilon) = \min(d, \; \epsilon/C)$,
so that $|\iota_1 - \iota_0| < \delta(\epsilon)$ implies $|\Delta(\iota_1)-\Delta(\iota_0)| < \epsilon$.
Thus, $\Delta(\iota)$ is continuous 
everywhere on $\mathcal{B}$.


\subsection{Continuity of $\bar\Delta(\iota)$}
\label{sec:continuity2}


Next, it is convenient to define a function $\bar{\Delta}(\iota)$ 
which is like $\Delta(\iota)$, except that it is non-infinite for any $\iota \in \mathbb{R}$, and its range is constrained to lie in $[-1,1]$:
\begin{align}
\bar{\Delta}(\iota) = 
\begin{cases}
1 &\mbox{if } \sigma \mbox{ is unbounded from above or if } \Delta(\iota)>1, \\
-1 &\mbox{if } \sigma \mbox{ is unbounded from below or if } \Delta(\iota)<-1, \\
\Delta(\iota) &\mbox{otherwise.} 
\end{cases}
\end{align}
The function $\bar\Delta(\iota)$ for the parameters
of figure \ref{fig:ODE}.a is shown in figure \ref{fig:ODE}.b.

We now prove that $\bar\Delta(\iota)$ is continuous at $\iota = \iota_0$ for all $\iota_0 \in \mathbb{R}$, considering three cases.
In the first case, consider $\iota_0$ for which (\ref{eq:PQ}) can be integrated to $\boozertor=2\pi$. Then due to the results of sections \ref{sec:bounded}-\ref{sec:continuity}, $\Delta(\iota)$ is non-infinite in a neighborhood of $\iota_0$. In this neighborhood, $\bar\Delta$ is a composition of continuous functions:  
$\bar{\Delta}(\iota) = \max(-1,\min(1,\Delta(\iota)))$,
hence $\bar{\Delta}(\iota)$ is continuous at this $\iota$.

In the second case, consider an $\iota_0$ for which the associated solution $\sigma_0$ is unbounded from above, so $\bar\Delta(\iota_0)=1$. This scenario can only happen if $\iota_0<0$. For any $\iota < \iota_0$, then $\sigma>\sigma_0$ by the monotonicity results
of section \ref{sec:uniqueness}, so $\sigma$ must diverge to $+\infty$, and
so $\bar{\Delta}(\iota)=1=\bar{\Delta}(\iota_0)$. To bound the behavior of $\bar\Delta$ when $\iota > \iota_0$, consider that since
$\sigma_0$ is unbounded from above, then for any quantity $q$, there must exist some $\boozertor_0 \in (0,2\pi)$ such that (\ref{eq:PQ}) (with $\iota_0$ and $\sigma_0$) can be integrated to $\boozertor_0$, and $\sigma_0(\boozertor_0) > q$. This statement is true in particular for the choice
\begin{align}
q = 2 + \sigma(0) + \int_0^{2\pi}d\boozertor |Q(\boozertor)|.
\end{align}
Since $\sigma$ must be a continuous function of $\iota$ at $\boozertor_0$ and $\iota_0$ (since the argument of section \ref{sec:continuity} applies at $\boozertor_0$ just as it does at $\boozertor=2\pi$), then there exists some $\delta$ such that for all $\iota$ satisfying $|\iota - \iota_0|<\delta$, then $|\sigma(\boozertor_0) - \sigma_0(\boozertor_0)| < 1$, so $\sigma(\boozertor_0) > q-1$. 
For such an $\iota$, if we require $|\iota - \iota_0|<|\iota_0|$
so $\iota <0$, then either $\sigma$ will diverge to $+\infty$ or else
\begin{align}
\Delta(\iota) 
=& -\sigma(0) + \sigma(\boozertor_0) + \int_{\boozertor_0}^{2\pi}d\boozertor \frac{d\sigma}{d\boozertor}
>-\sigma(0) + q-1 + \int_{\boozertor_0}^{2\pi}d\boozertor \frac{d\sigma}{d\boozertor} \\
=& 1 + \int_0^{2\pi}d\boozertor|Q(\boozertor)| + \int_{\boozertor_0}^{2\pi}d\boozertor [-\iota(P+\sigma^2)-Q]
\ge 1+\int_0^{2\pi}d\boozertor|Q(\boozertor)| - \int_{\boozertor_0}^{2\pi}d\boozertor Q
\ge 1. \nonumber
\end{align}
Therefore, as long as $|\iota - \iota_0| < \min(\delta,|\iota_0|)$, 
then $\bar\Delta(\iota)=1$, so $|\bar\Delta(\iota) - \bar\Delta(\iota_0)|<\epsilon$ for any $\epsilon>0$.
Therefore $\bar\Delta(\iota)$ is continuous at $\iota = \iota_0$.

For the third case, in which $\iota_0$ is such that $\sigma_0$ diverges to $-\infty$, continuity may be proved analogously to case 2, with a few appropriate changes of sign.


\subsection{Existence of a solution}

Finally, we can prove that at least one value of $\iota$ exists for which the solution $\sigma$ of (\ref{eq:PQ}) is periodic.
Let $\bar P=\int_0^{2\pi}P\, d\boozertor$ and $\bar Q=\int_0^{2\pi}Q\, d\boozertor$, and 
let 
\begin{equation}
\iota_n = \min\left(0, \; 
-\bar{Q}/\bar{P}
\right).
\end{equation}
Since $\iota_n \le 0$, either the associated $\sigma$ diverges to $+\infty$ or else we can integrate (\ref{eq:PQ}) over $[0,\,2\pi]$ to obtain
\begin{equation}
\Delta(\iota_n) \ge 
-\iota_n \bar{P}-\bar{Q} \ge 0.
\end{equation}
Thus, $\bar\Delta(\iota_n) \ge 0$.
Similarly, let 
$\iota_p = \max\left(0, 
-\bar{Q}/\bar{P}
\right)$. 
Since $\iota_p \ge 0$, either the associated $\sigma$  diverges to $-\infty$ or else we can integrate over $[0,\,2\pi]$ to obtain
\begin{equation}
\Delta(\iota_p) \le 
-\iota_p \bar{P} - \bar{Q} \le 0.
\end{equation}
Then $\bar\Delta(\iota_p) \le 0$. We have thus shown that values of $\iota$ exist for which $\bar\Delta(\iota)$ is non-positive and non-negative,
and we have shown $\bar\Delta(\iota)$ is continuous. By the intermediate value theorem, then there must exist
an $\iota$ in the interval $[\iota_n, \, \iota_p]$ for which $\bar\Delta(\iota)=0$. Therefore $\Delta(\iota)=0$, corresponding to a periodic $\sigma$.
Thus, it is guaranteed that precisely one periodic solution $\{\iota,\,\sigma(\boozertor)\}$ of (\ref{eq:PQ}) exists.

\bibliographystyle{jpp}

\bibliography{quasisymmetry_cylindrical_II}

\end{document}